\documentclass[11pt]{article}

\usepackage{aas_macros,amsmath,amssymb,comment,cite,esint,graphicx,mathtools}
\usepackage[margin=.8in,letterpaper]{geometry}
\usepackage[colorlinks=true]{hyperref}
\usepackage[affil-it]{authblk}
\usepackage{subcaption}
\usepackage[utf8]{inputenc}
\usepackage{mathrsfs}
\usepackage{appendix}
\usepackage{amssymb}
\usepackage{float}
\usepackage{color}
\usepackage{cite}
\usepackage{hyperref}
\hypersetup{pageanchor=false}
\usepackage{indentfirst}
\usepackage{url}
\usepackage{float}
\usepackage{caption}
\usepackage[numbers,square,comma,sort&compress,merge]{natbib}
\usepackage{esint}
\usepackage{overpic}
\usepackage{graphicx}
\usepackage{epsf,amsmath,bbold,amsfonts,stmaryrd}
\usepackage{textcomp}
\usepackage{ulem}
\usepackage{tikz}

\numberwithin{equation}{section}
\let\originalleft\left
\let\originalright\right
\renewcommand{\left}{\mathopen{}\mathclose\bgroup\originalleft}
\renewcommand{\right}{\aftergroup\egroup\originalright}
\mathcode`\*="8000
{\catcode`\*=\active\gdef*{\mathclose{}\,\mathopen{}}}

\newcommand{\bea}{\setlength\arraycolsep{2pt} \begin{eqnarray}}
\newcommand{\eea}{\end{eqnarray}}
\newcommand{\nn}{\nonumber}
\newcommand{\mo}{\mathcal{O}}
\newcommand{\df}{\mathrm{d}}

\def\nn{\nonumber}

\def\bag{\begin{aligned}}
\def\eag{\end{aligned}}
\def\bea{\begin{eqnarray}}
\def\eea{\end{eqnarray}}
\def\ba{\begin{array}}
\def\ea{\end{array}}

\def\bc{\begin{center}}
\def\ec{\end{center}}

\begin{document}
\title{Revisiting the shadow of Johannsen-Psaltis black holes}
	
\author{Xinyu Wang$^{1, 2}$, Zhixing Zhao$^{1, 2}$, Xiao-Xiong Zeng$^{3, 4\ast}$, Xin-Yang Wang$^{2, 5\ast}$}
\date{}
	
\maketitle
\vspace{-10mm}

\begin{center}
{\it
$^1$ School of Physics and Astronomy, Beijing Normal University,
Beijing 100875, China\\\vspace{4mm}

$^2$Key Laboratory of Multiscale Spin Physics (Ministry of Education), Beijing Normal University, Beijing 100875, China\\\vspace{4mm}

$^3$ State Key Laboratory of Mountain Bridge and Tunnel Engineering,
Chongqing Jiaotong University, Chongqing 400074, China\\\vspace{4mm}

$^4$ Department of Mechanics, Chongqing Jiaotong University, Chongqing 400074, China\\\vspace{4mm}

$^5$ Faculty of Arts and Sciences, Beijing Normal University, Zhuhai 519087, China\\\vspace{4mm}
}
\end{center}

\vspace{8mm}

\begin{abstract}
    The Johannsen-Psaltis (JP) metric provides a robust framework for testing the “no-hair theorem” of astrophysical black holes due to the regular spacetime configuration around JP black holes. Verification of this theorem through electromagnetic spectra often involves analyzing the photon sphere near black holes, which is intrinsically linked to black hole shadows. Investigating the shadow of JP black holes offers an effective approach for assessing the validity of theorem. Since the Hamilton-Jacobi equation in the JP metric does not permit exact variable separation, an approximate analytical approach is employed to calculate the shadow, while the backward ray-tracing numerical method serves as a rigorous alternative. For JP black holes with closed event horizons, the approximate analytical approach reliably reproduces results obtained through the numerical computation. However, significant discrepancies emerge for black holes with non-closed event horizons. As the deviation parameter $\epsilon_3$ increases, certain regions of the critical curve transition into non-smooth configurations. Analysis of photon trajectories in these regions reveals chaotic dynamics, which accounts for the failure of the approximate analytical method to accurately describe the shadows of JP black holes with non-closed event horizons.
\end{abstract}

\vfill{\footnotesize \flushleft $\ast$ Corresponding author: \\
xxzengphysics@163.com\\
xinyangwang@bnu.edu.cn
}

\maketitle

\newpage
\baselineskip 18pt
\section{Introduction}\label{sec1}

Black holes represent a unique spacetime structure predicted by General Relativity (GR). The outer boundary of black holes is enclosed by a null hypersurface known as the event horizon. Numerous significant properties of black holes are revealed through the characteristics of the event horizon. Initially, the study of black holes was largely theoretical, as their existence lacked experimental confirmation. However, with advancements in technology and improvements in detection techniques, numerous experiments have provided both direct and indirect evidence supporting the existence of black holes in the observable universe. Prominent milestones of these advancements include the detection of gravitational waves from black hole mergers by the LIGO and Virgo collaborations and the imaging of black hole shadows by the Event Horizon Telescope (EHT) \cite{EventHorizonTelescope:2019dse,EventHorizonTelescope:2019uob,EventHorizonTelescope:2019jan,EventHorizonTelescope:2019ths,EventHorizonTelescope:2019pgp,EventHorizonTelescope:2019ggy,EventHorizonTelescope:2021bee}. The shadow curves of Schwarzschild and Kerr black holes are formed by null trajectories from photons with critical impact parameters in the strong light-bending region \cite{1983mtbh.book.....C,Cunha:2018acu,Perlick:2021aok,Wang:2023nwd,Wang:2024uda}. These shadows manifest as luminous halos in the image, composed of photons from the hot gas surrounding the black hole that are bent before reaching the telescope. The black hole image not only provides direct evidence of the existence of black holes but also serves as a tool to study other properties of black holes. The advent of EHT has rekindled interest in calculating black hole shadows for a wide range of solutions of gravitational theories, including those in modified gravity \cite{Wei:2013kza,Grenzebach:2014fha,Cunha:2016bpi,Cunha:2016wzk,Guo:2020zmf,Chen:2023trn,Hou:2021okc,Wei:2020ght,Ovgun:2020gjz,Kuang:2022ojj} and scenarios involving new physics \cite{Haroon:2018ryd,Konoplya:2019sns,Chen:2019fsq,Chen:2021lvo,Faraji:2024ein,Zhang:2024hjr, Zeng:2024ptv} etc.

The ``no-hair theorem'', a fundamental principle of black hole physics in GR, asserts that stationary black holes in Einstein-Maxwell gravity are fully characterized by three observable parameters: mass, charge, and angular momentum \cite{newman1965metric}. According to this theorem, all additional information about the matter fields that formed black holes or crossed the event horizon becomes permanently inaccessible to external observers once the black holes reach equilibrium through gravitational wave and electromagnetic radiation emission. Observational evidence supports the existence of astrophysical black holes, typically defined by the mass and angular momentum. The Kerr solution, consistent with the no-hair theorem, serves as a reasonable model for describing astrophysical black holes. However, astrophysical black holes commonly reside in non-vacuum environments containing other celestial objects, dark matter, and matter fields, which can induce deviations from the Kerr metric and alter distinct features of the Kerr solution \cite{Ryan:1997hg, Gair:2011ym}. These deviations primarily stem from differences in mass and angular momentum, the fundamental parameters of the no-hair theorem, leaving its strict validity for astrophysical black holes unverified \cite{Johannsen:2010xs, Johannsen:2010ru, Johannsen:2010bi}. Therefore, it is crucial to develop a robust metric that systematically accounts for deviations from the Kerr metric to evaluate whether astrophysical black holes conform to the no-hair theorem \cite{Collins:2004ex, Vigeland:2009pr,Glampedakis:2005cf,Mankot1992GeneralizationsOT,Vigeland:2011ji}. Many proposed spacetime metrics that deviate from the Kerr solution introduce irregularities, such as singularities or closed timelike curves, outside the event horizon \cite{Johannsen:2013szh}. These irregularities obstruct the use of photon circular orbits and the innermost stable circular orbit (ISCO), which are widely regarded as significant indicators in electromagnetic spectrum observations for testing the no-hair theorem \cite{Johannsen:2013rqa}. Consequently, effective testing of the no-hair theorem for astrophysical black holes necessitates an alternative metric that deviates from the Kerr metric while avoiding singularities or closed timelike curves outside the event horizon.

Johannsen and Psaltis \cite{Johannsen:2011dh} developed a set of of Kerr-like metrics, known as the Johannsen-Psaltis (JP) metric, which characterizes stationary, axisymmetric, and asymptotically flat vacuum spacetime. This metric allows black holes to exhibit angular momentum across the entire allowable range and avoids spacetime irregularities, such as singularities or closed timelike curves, outside the event horizon, thereby ensuring regularity in the external spacetime. The JP metric is characterized by a set of parameters that quantify deviations from the Kerr metric, seamlessly reducing to the Kerr metric as these parameters approach zero. Unlike the Kerr metric, which serves as a vacuum solution to the Einstein field equations, the JP metric arises as a specific solution within a modified theory of gravity. The investigation of the no-hair theorem primarily involves analyzing the spacetime structure described by the JP metric, irrespective of its origin. Therefore, explicitly specifying the gravitational field equations in the corresponding modified theory of gravity is not required for this analysis. It is important to highlight that the JP metric retains compatibility with the backward ray-tracing method used for calculating photon trajectories in Kerr spacetime, as its non-zero metric components correspond closely to those of the Kerr metric. This consistency ensures the applicability of backward ray-tracing techniques for interpreting observational phenomena, such as black hole shadows and accretion dynamics, through electromagnetic spectra. As a result, the shadow of a JP black hole provides a robust framework for testing the no-hair theorem within the context of electromagnetic spectrum analysis.

The characteristics of spacetime described by JP metric, along with the shadow and accretion phenomena of JP black holes, have been systematically investigated. The property of the event horizon of JP black holes and the anomalous regions of spacetime described by JP metric have been extensively analyzed by Johannsen \textit{et al.} \cite{Johannsen:2013rqa}. The shadow of JP black holes has been analytically calculated \cite{Atamurotov:2013sca,Bambhaniya:2021ybs,Khodadi:2021gbc}, and hot spot models in the JP spacetime have also been explored \cite{Li:2014fza,Liu:2014awa}. Building on these findings, the accretion phenomena of JP black holes have been investigated, and the images of thin and thick accretion disks around JP black holes are studied separately \cite{Bambi:2012tg, Kong:2014wha, Li:2012ra}. Specifically, the Bondi accretion in the accretion phenomenon of black holes was investigated by John and Stevens \cite{John:2019rhj}. The JP metric has been generalized to incorporate charge by Rahim and Saifullah \cite{Rahim:2018ruj}, and this extended metric is referred as the charged version of JP metric. Based on the extended metric, the circular photon orbits and ISCOs were computed employing methods analogous to those used in the JP metric spacetime. 

The computation of black hole shadows primarily concerns the propagation of photons in spacetime, and the trajectories of photons are predominantly determined by the Hamilton-Jacobi equation. In the spacetime described by the Kerr metric, three conserved quantities can be identified: the energy $E$ and angular momentum $L$, derived from the Killing vector fields associated with the spacetime symmetries, and the mass of the particle $m$. Furthermore, a fourth conserved quantity referred as the Carter constant arises from a rank-2 Killing tensor contracting with two Killing vectors \cite{Carter:1968rr}. The four conserved quantities reduce the equations of motion to a first-order form, with variable separability ensured by the inclusion of the Carter constant. This separability facilitates the computation of Kerr black hole shadows through fully analytical methods. Moreover, the Kerr metric is classified as a Petrov-type D metric in the Petrov classification because the Carter constant is well-defined in this spacetime. However, the Carter constant cannot be defined in the spacetime described by the JP metric, as this metric belongs to the Petrov-type I classification. It implies that the Hamilton-Jacobi equation is not separable, rendering the computation of a JP black hole shadow through analytical methods infeasible \cite{Johannsen:2013szh,Johannsen:2013vgc,Glampedakis:2018blj,Younsi:2021dxe}. Therefore, the shadow of a JP black hole can only be calculated using the approximate analytical method or the exact numerical method.

The structure of the paper is as follows. In Sec. \ref{Metric and Separability}, we review the basic concepts and properties of the JP metric and JP black holes. Section \ref{investigationjpshadow} investigates the shadow of JP black holes through the application of both the approximate analytical approach and the numerical computational method. In Sec. \ref{Geodesic Equations}, we discuss the approximate separation of variables for the Hamilton-Jacobi equation and derive the equations of motion of photons in the JP spacetime. Sec. \ref{Analytical Calculation} introduces the theoretical framework for calculating black hole shadows using the approximate analytical and the numerical approach. In Sec. \ref{Numerical Calculation of the Image and Shadow}, we present and compare the results of JP black hole shadows obtained through the approximate analytical and numerical methods. The paper ends with conclusions in Sec. \ref{Summary}.

In this work, we have set the fundamental constants $c$, $G$ to unity, and we will work in the convention $(-,+,+,+)$.
	
\section{A brief review of the JP metric and JP black holes}\label{Metric and Separability}

The JP metric, incorporating deviations from the Kerr metric, serves as a framework for investigating the applicability of the no-hair theorem to astrophysical black holes. The specific form of the metric in Boyer-Lindquist (BL) coordinates is expressed as \cite{Johannsen:2011dh} 
\bea\label{metric}
\df s^2&=& -H(r,\theta)\left(1-\frac{2 M r}{\Sigma}\right) d t^2-\frac{4 a M r \sin ^2 \theta}{\Sigma}H(r,\theta) d t d \phi+\frac{\Sigma H(r,\theta)}{\Delta+h(r, \theta)a^2 \sin ^2 \theta } d r^2 \nn\\
&& +\Sigma d \theta^2+\sin ^2\theta\left[\Sigma+H(r, \theta) \frac{a^2(\Sigma+2 M r) \sin ^2 \theta}{\Sigma}\right] d \phi^2\,,
\eea
where
\bea
\Delta=r^2-2 M r+a^2, \quad \Sigma=r^2+a^2 \cos ^2 \theta\,.
\eea
In the metric expression, the parameter $M$ represents the mass of the black hole, while $a = J / M$ denotes the angular momentum per unit mass, with $J$ being the angular momentum of the black hole. The function $H (r\,, \theta)$ in the JP metric is expressed as 
\begin{equation}\label{functionh}
    H(r,\theta)=1 + h (r, \theta)\,,
\end{equation}
where the function $h \left(r\,, \theta \right)$ is defined as 
\begin{equation}\label{functionhrtheta}
    \begin{split}
        h (r, \theta) \equiv \sum_{k=0}^{\infty}\left(\epsilon_{2 k}+\epsilon_{2 k+1} \frac{M r}{\Sigma} \right) \left(\frac{M^2}{\Sigma} \right)^k\,.
    \end{split}
\end{equation}
The JP metric, like the Kerr metric, is stationary and axisymmetric, describing an asymptotically flat spacetime. The primary distinction between the JP metric and the Kerr metric lies in the inclusion of the function $H (r\,, \theta)$, which introduces deviations from the standard Kerr metric. As shown in Eq. (\ref{functionhrtheta}), the series of parameters $\epsilon_{k} \, \left(k = 0\,, \cdots \infty \right)$ embedded in $h (r\,, \theta)$ quantifies these deviations. The JP metric reduces to the Kerr metric when all the parameters in the series are set to zero and further simplifies to the Schwarzschild metric when the spin parameter $a$ is set to zero. The inclusion of $H(r, \theta)$ introduces novel properties absent in the Kerr metric, while preserving essential characteristics of the Kerr metric, as the Kerr metric serves as a limiting case of the JP metric. These retained properties are crucial for testing the no-hair theorem. It is important to note that the JP metric represents a family of solutions derived from an unspecified set of vacuum field equations, rather than directly from the Einstein field equations, with the explicit form of these vacuum field equations remaining unknown. However, testing the no-hair theorem through electromagnetic spectrum analysis requires only an appropriate spacetime framework for investigating particle motion, provided that the chosen spacetime satisfies the conditions of Einstein equivalence principle.

Before analyzing the motion of particles in the spacetime described by the JP metric, it is essential to clarify the constraints on the function $H (r\,, \theta)$ in the metric. This function should satisfy additional conditions to ensure that the JP spacetime remains asymptotically flat and accurately reflects observed deviations from the Kerr metric in the weak-field regime. In the gravitational theories encompassed by GR, stationary and asymptotically flat spacetimes can be approximated by a Schwarzschild-like metric at sufficiently large distances from the gravitational source. This concept also extends to more general spacetime corresponding to solutions of non-Einstein field equations. Therefore, when the coordinate $r$ satisfies the conditions $r \gg M$ and $r \gg a$, the JP metric should assume an asymptotically Schwarzschild-like form. In this regime, all components corresponding to the Schwarzschild part of this asymptotically Schwarzschild-like metric are well-defined and are proportional to $1 / r^n$, where $n = 0$ and $n = 1$. This requirement implies that the terms in the metric associated with the function $h (r\,, \theta)$ should satisfy $\mathcal{O} (1 / r^n)\,, n \ge 2$, and the first two parameters $\epsilon_0$ and $\epsilon_1$ must vanish in the asymptotic form of the JP metric. Furthermore, two additional parameters, $\beta$ and $\gamma$, can be introduced to express the parameter $\epsilon_2$ in the function $h (r\,, \theta)$ as
\begin{equation}
    \begin{split}
        \epsilon_2 = 2 \left(\beta - 1 \right)\,, \qquad \gamma = 1\,.
    \end{split}
\end{equation}
When $\beta = \gamma = 1$, the asymptotic form of the JP metric reduces to the standard Schwarzschild metric. Based on the Lunar Laser Ranging Experiment, the range of the parameter $\beta$ is constrained as $\left|\beta - 1 \right| \le 2.3 \times 10^{-4}$, while the range of the parameter $\epsilon_2$ in the function $h (r\,, \theta)$ is bounded by $\left|\epsilon_2 \right| \le 4.6 \times 10^{-4}$. Since the value of the parameter $\epsilon_2$ is sufficiently small, its contribution to the asymptotic form of the JP metric is negligible. Therefore, the parameter $\epsilon_2$ is set to zero, i.e., $\epsilon_2 = 0$. Furthermore, we assume $\epsilon_k = 0$ for $k > 3$, retaining only $\epsilon_3$ in the expression of $h (r\,, \theta)$. Under these conditions, the function $h (r\,, \theta)$ can be simplified as \cite{Johannsen:2011dh}
\begin{equation}\label{hrthetae3}
    \begin{split}
        h(r, \theta) = \epsilon_3 \frac{M^3 r}{\Sigma^2}\,.
    \end{split}
\end{equation}
This choice is motivated by the fact that the parameter $\epsilon_3$ is not constrained by current observational or experimental results within GR. Retaining this parameter in the metric allows for a parametric investigation into the characteristics of strong gravitational fields.

The JP black hole exists within the spacetime described by the JP metric when an event horizon is present. The location of the event horizon for the JP black hole is determined by the following equation
\begin{equation}\label{governingeqeh}
    \begin{split}
        g_{t \phi}^2 - g_{tt} g_{\phi \phi} = 0\,.
    \end{split}
\end{equation}
The JP black hole, like the Kerr black hole, can simultaneously exhibit both an inner and an outer horizon, as the governing equation in Eq. (\ref{governingeqeh}) allows for multiple solutions. However, the inclusion of the parameter $\epsilon_3$ introduces unique characteristics to the event horizon of the JP black hole that are absent in the Kerr metric. Since the critical connection between the properties of the event horizon of JP black holes and its shadow, a detailed analysis of these distinctive features is warranted. It is also important to note that the spin parameter $a$ for Kerr black holes is constrained to the range $0 \leq a \leq 1$, whereas the parameter $\epsilon_3$ in the JP metric allows the spin parameter of JP black holes to exceed this range, i.e., $a > 1$, leading to the formation of superspinning black holes. Since Kerr black holes are not defined for $a > 1$, the spin parameter is restricted to $0 \leq a \leq 1$ in the subsequent analysis to ensure a consistent comparison between JP and Kerr black holes.

\begin{figure}[htbp]
    \centering
    \begin{minipage}[b]{0.46\textwidth}
        \centering
        \includegraphics[width=\textwidth]{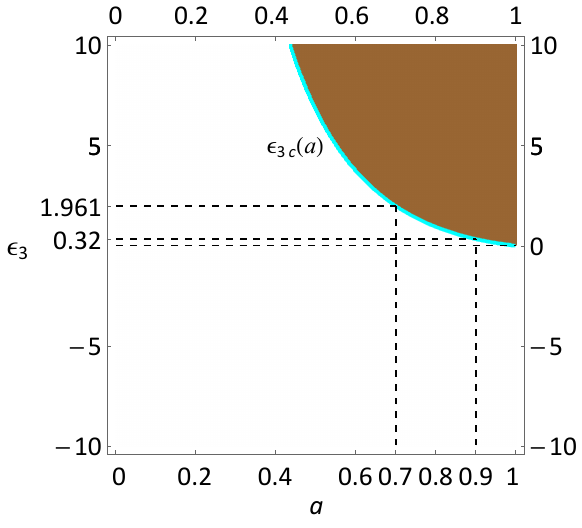}
        \caption*{(a)}
    \end{minipage}
    \begin{minipage}[b]{0.46\textwidth}
        \centering
        \includegraphics[width=\textwidth]{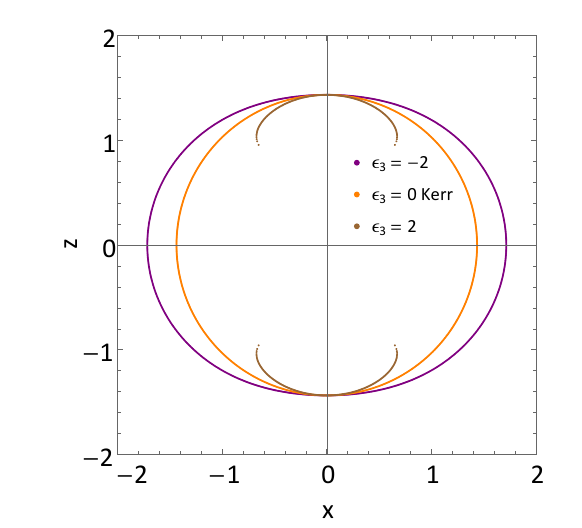}
        \caption*{(b)}
    \end{minipage}

    \vspace{1em}

    \begin{minipage}[b]{0.45\textwidth}
        \centering
        \includegraphics[width=\textwidth]{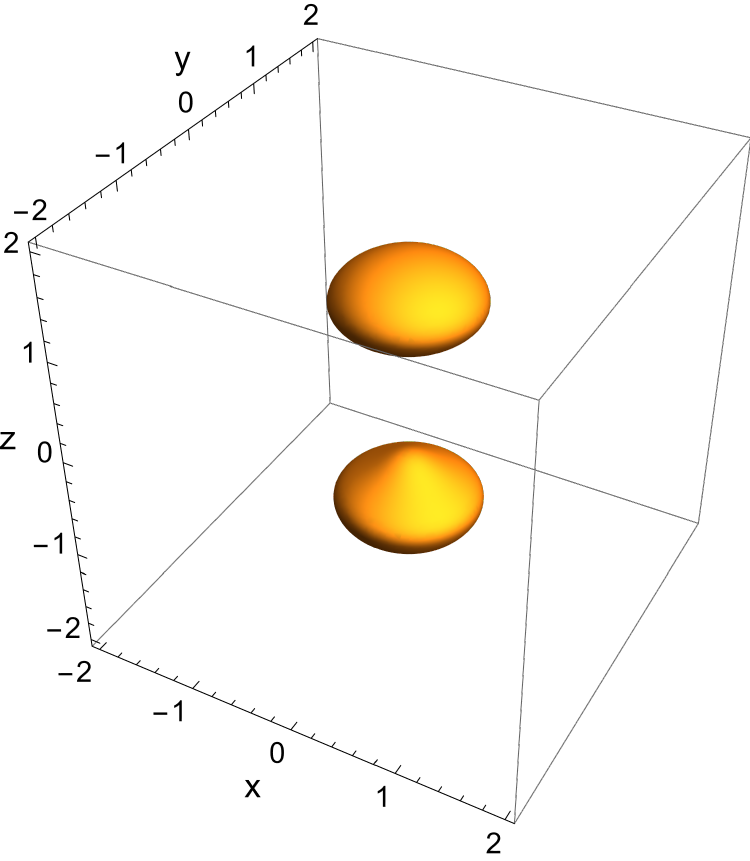}
        \caption*{(c)}
    \end{minipage}
    \begin{minipage}[b]{0.45\textwidth}
        \centering
        \includegraphics[width=\textwidth]{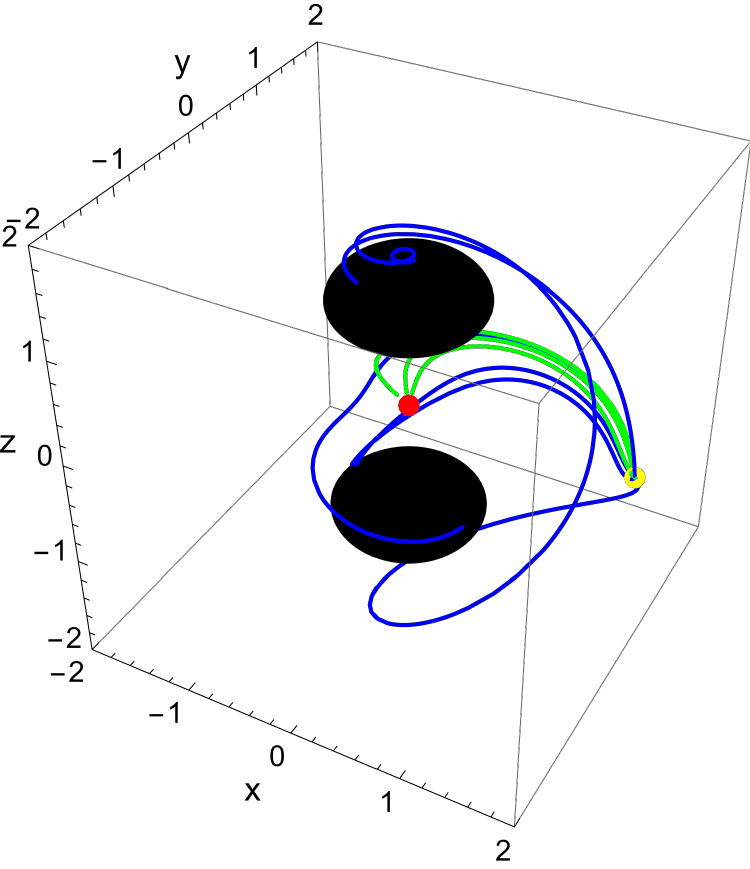}
        \caption*{(d)}
    \end{minipage}
    \caption{(a) The parameter space distribution of $\epsilon_3$ and $a$ is constructed to analyze the presence of an event horizon for $a \leq 1$. (b) The cross-sectional projections of the event horizon onto the $xz$-plane for $a = 0.9$ and $\epsilon_3 = -2, 0, 2$ are calculated to illustrate the influence of the deviation parameter $\epsilon_3$ on the shape of the event horizon. (c) The event horizon is depicted in the three-dimensional $xyz$-coordinates for the parameters $a = 0.9$ and $\epsilon_3 = 2$. (d) Several null geodesics emanating from a yellow point located outside the naked singularity are illustrated for the parameters $a = 0.9$ and $\epsilon_3 = 2$.}
    \label{horizonall}
\end{figure}
The relationship between the event horizon of the JP black hole, the spin parameter $a$, and the deviation parameter $\epsilon_3$ is depicted in Fig. \ref{horizonall} (a). In this figure, a critical curve, $\epsilon_{3c} (a)$, divides the two-dimensional parameter space into white and brown regions. The white region corresponds to combinations of the spin $a$ and the deviation parameter $\epsilon_3$ that result in a closed event horizon, signifying the presence of a JP black hole in spacetime. Conversely, the brown region represents parameter values for which the event horizon described by the JP metric becomes open, indicating the emergence of a naked singularity coexisting with a JP black hole within this parameter range. Specifically, two specific points on the critical curve, $\epsilon_{3c}(0.9) = 0.32$ and $\epsilon_{3c}(0.7) = 1.96$, are highlighted for further analysis and discussion. Fig. \ref{horizonall} (b) illustrates the projection of the event horizon of the JP black hole onto the $xz$-plane for $a = 0.9$ and $\epsilon_3 = -2\,, 0\,, 2$. When $\epsilon_3 = 0$, the JP metric reduces to the Kerr metric, and the event horizon corresponds exactly to that of a Kerr black hole. For $\epsilon_3 < 0$, the event horizon of the JP black hole remains closed and adopts a more oblate elliptical shape compared to the event horizon of the Kerr black hole. Conversely, for $\epsilon_3 > 0$, the event horizon becomes non-closed, leading to the emergence of a naked singularity in the spacetime. Fig. \ref{horizonall} (c) presents a three-dimensional visualization of the non-closed event horizon for $a = 0.9$ and $\epsilon_3 = 2$. The event horizon in this scenario exhibits a distinctive dumbbell-like configuration, highlighting the unique geometric properties associated with the non-closed event horizon in the spacetime described by the JP metric. 

In scenarios where spacetime features a non-closed event horizon, the presence of a naked singularity may lead to causality violations. To analyze the shadow of a JP black hole with a non-closed event horizon, it is crucial to ensure that photons originating near the naked singularity cannot reach distant observers. This verification is essential to maintain consistency with the principles of causality in the resulting JP black hole shadow. As depicted in Fig. \ref{horizonall} (d), the numerical approach utilizing the backward ray-tracing method is applied to analyze several null geodesics originating from the yellow point located outside the non-closed event horizon of JP black holes. The results show that photons confined to the equatorial plane ultimately fall into the naked singularity, while most photons originating outside the equatorial plane follow null geodesics that intersect the event horizon, with some trajectories approaching the naked singularity. Further analysis shows that when the yellow point is sufficiently distant from the event horizon and the coordinate time in its reference frame is used for measurement, the time required for photons traveling along null geodesics to approach the singularity diverges asymptotically to infinity. This implies that distant observers cannot directly detect the naked singularity, and all photons received by the observer adhere to the principles of causality. These findings suggest that investigating the shadow of a JP black hole is theoretically feasible even in the presence of a naked singularity. In the subsequent analysis, both analytical and numerical approaches will be employed to compute and examine the shadows of JP black holes with closed and non-closed event horizons. A detailed comparative evaluation will be performed to assess the consistency and discrepancies between the results derived from these two methodologies. For simplicity and without loss of generality, the value of the parameter $M$ is set to $M = 1$ in the following discussion.

\section{Investigations of JP black hole shadow}\label{investigationjpshadow}

Based on the preceding analysis of the JP metric, the function $H(r, \theta)$ introduces deviations from the Kerr metric. As shown in Eqs. (\ref{functionh}) and (\ref{hrthetae3}), the function $H(r, \theta)$ primarily depends on the parameter $\epsilon_3$ in the function $h(r, \theta)$. When the value of $\epsilon_3$ is set to zero, the function $H(r, \theta)$ vanishes from the expression of the JP metric, and the JP metric smoothly reduces to the Kerr metric. The absence of the Carter constant in the spacetime described by the JP metric precludes variable separation in the Hamilton-Jacobi equation for photons, rendering it impossible to compute the shadow of JP black holes through a fully analytical approach. Conversely, the Hamilton-Jacobi equation of photons in the Kerr metric spacetime allows complete variable separation. Although the JP metric deviates from the Kerr metric due to the presence of the parameter $\epsilon_3$, it smoothly reduces to the Kerr metric as the value of $\epsilon_3$ approaches zero. This suggests that the Hamilton-Jacobi equation in the JP metric spacetime can approximately allow variable separation under the limiting condition $\epsilon_3 \to 0$. In the subsequent analysis, the Hamilton-Jacobi equation governing photon motion, allowing for approximate variable separation under the limiting condition $\epsilon_3 \to 0$, will be established. The specific equations of motion for photons in the spacetime described by the JP metric will be derived, and the shadow of the JP black hole will be calculated through an approximate analytical method. In parallel, the shadow of the JP black hole will be calculated via the rigorous numerical computation. A comparative analysis of the approximate analytical results and the exact numerical findings will be performed to evaluate the applicability and limitations of the approximate analytical approach in studying the shadow of JP black holes.

\subsection{Approximate analytical derivation of motion equations}
\label{Geodesic Equations}

\begin{figure}[htbp]
    \centering
    \begin{minipage}[b]{0.45\textwidth}
        \centering
        \includegraphics[width=\textwidth]{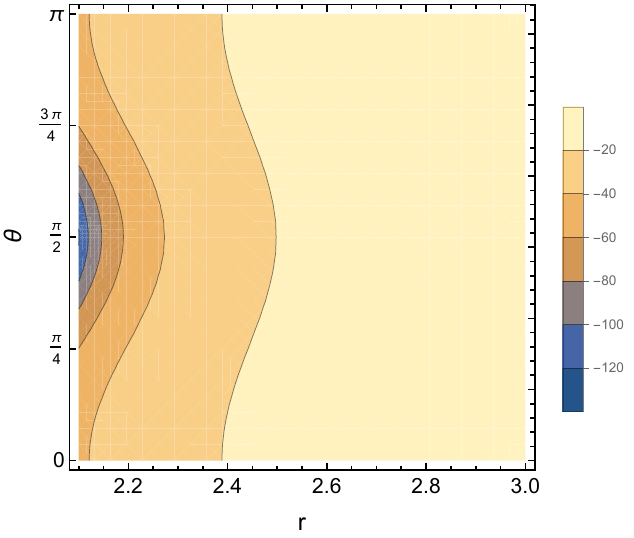}
        \caption*{(a) $a = 0.9\,, \epsilon_3 = - 8$} 
    \end{minipage}
    \begin{minipage}[b]{0.45\textwidth}
        \centering
        \includegraphics[width=\textwidth]{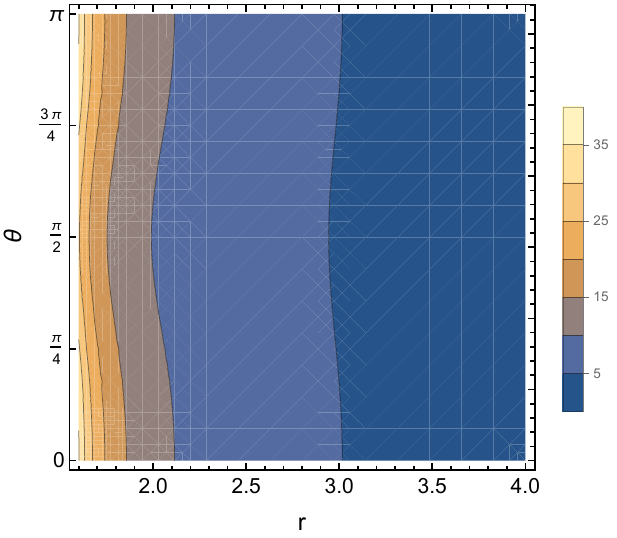}
        \caption*{(b) $a = 0.9\,, \epsilon_3 = 8$}
    \end{minipage}
    \caption{The correlation between the function $\mathcal{I} \left(r\,, \theta \right)$ and the radial coordinate $r$, as well as the angular coordinate $\theta$. (a) The spacetime of JP metric with $a = 0.9$ and $\epsilon_3 = - 8$. (b) The spacetime of JP metric with $a = 0.9$ and $\epsilon_3 = 8$.}
    \label{relationrandtheta}
\end{figure}

As previously mentioned, the Kerr metric is classified as Petrov-type D within the Petrov classification scheme. The Carter constant, defined by the Killing vector fields in the spacetime of the Kerr metric \cite{Carter:1968rr}, allows for the separation of variables in the Hamilton-Jacobi equation. This separability facilitates the computation of null geodesics and black hole shadows in Kerr spacetime. In contrast, the JP metric is categorized as the Petrov-type I classification, which precludes the definition of a Carter constant in JP spacetime \cite{Glampedakis:2018blj,Johannsen:2013vgc,Johannsen:2013szh,Younsi:2021dxe}. Consequently, conventional methods for calculating black hole shadows that rely on variable separation in the equations of motion are not directly applicable to JP black holes \cite{1983mtbh.book.....C}. However, as the JP metric transitions to the Kerr metric under the limiting condition $\epsilon_3 \to 0$, the approximate separability of variables in the Hamilton-Jacobi equation for particle motion becomes achievable. This limiting condition will be employed to derive approximate equations of motion for particles in JP spacetime from the Hamilton-Jacobi equation. The Hamilton-Jacobi equation for a general system is expressed as
\begin{equation}\label{HJ}
    \begin{split}
        \frac{\partial S}{\partial \lambda} + H = 0\,,
    \end{split}
\end{equation}
where $S$ denotes the action of the system, $\lambda$ represents an affine parameter, and $H$ is the Hamiltonian function describing the system. For a free particle undergoing geodesic motion in a spacetime, the Hamiltonian function in the Hamilton-Jacobi equation can be expressed as
\begin{equation}\label{hamiltonexp}
    \begin{split}
        H = \frac{1}{2} g^{\mu \nu} \frac{\partial S}{\partial x^\mu} \frac{\partial S}{\partial x^\nu} = \frac{1}{2} g^{\mu \nu} p_{\mu} p_{\nu} = - \frac{1}{2} m^2\,,
    \end{split}
\end{equation}
and the four momentum of the particle $p_{\mu}$ is defined as 
\begin{equation}\label{deffourmomentum}
    \begin{split}
        p_{\mu} = \frac{\partial S}{\partial x^{\mu}}\,.
    \end{split}
\end{equation}
The parameter $m$ in Eq. (\ref{hamiltonexp}) denotes the mass of the particle. As the JP metric converges to the Kerr metric under the limiting condition $\epsilon_3 \to 0$, the Hamilton-Jacobi equation describing particle geodesics approximately permits variable separation. When variable separation is applicable, the action $S$ can be expressed as
\bea\label{formalexpaction}
S=\frac{1}{2} m^2 \lambda-\mathcal{E} t+\mathcal{L} \phi+S_r(r)+S_\theta(\theta)\,,
\eea
where $m = 0$ corresponds to massless particles, and $m = 1$ represents massive particles. The quantities $\mathcal{E}$ and $\mathcal{L}$ on the right-hand side of Eq. (\ref{formalexpaction}) denote the energy and angular momentum in the spacetime described by the JP metric. These quantities are conserved due to the stationary and axisymmetric nature of the JP metric and are associated with the Killing vectors $\partial_t$ and $\partial_\phi$ in the spacetime. The terms $S_r (r)$ and $S_\theta (\theta)$ in Eq. (\ref{formalexpaction}) represent the contributions to the action associated with the radial coordinate $r$ and the angular coordinate $\theta$, respectively. Substituting the explicit form of the JP metric from Eq. (\ref{metric}) into the Hamilton-Jacobi equation in Eq. (\ref{HJ}), the equations of motion for massless or massive particles in this spacetime can be expressed as 
\bea\label{hamiltonjacobieq}
    \begin{aligned}
        &
        \left\{\frac{\left(\Delta+h a^2 \sin ^2 \theta\right)^2}{\Delta(1+h)}\left(\frac{\partial S_r}{\partial r}\right)^2-\frac{\left[\left(r^2+a^2\right) \mathcal{E}-a \mathcal{L}\right]^2}{\Delta}+\left[(\mathcal{L}-a \mathcal{E})^2+m^2 r^2\right] + \mathcal{I} \left(r\,, \theta \right) \right\}+\\
        &\left\{\frac{\left(\Delta+h a^2 \sin ^2 \theta\right)}{\Delta}\left(\frac{\partial S_\theta}{\partial \theta}\right)^2+\left(a^2 \cos ^2 \theta+\frac{a^2 \sin ^2 \theta h}{\Delta} \Sigma\right) m^2 + \left(\mathcal{L}^2 \csc ^2 \theta-a^2 \mathcal{E}^2\right) \cos ^2 \theta\right\}=0\,,
    \end{aligned}
\eea
where the specific expression for the function $\mathcal{I} \left(r\,, \theta \right)$ within the first set of curly brackets in the Hamilton-Jacobi equation can be given as
\begin{equation}
    \begin{split}
        \mathcal{I} \left(r\,, \theta \right) = \frac{\mathcal{E}^2 h \Sigma^2}{\Delta(1+h)}\,.
    \end{split}
\end{equation}
For the Hamilton-Jacobi equation of particles in Eq. (\ref{hamiltonjacobieq}), the terms within the first set of braces describe the evolution of the radial coordinate $r$ along the geodesic trajectory of particles, while the terms within the second set of braces characterize the evolution of the angular coordinate $\theta$ as the particles propagates through spacetime. The explicit form of the Hamilton-Jacobi equation governing particles motion in the spacetime described by the JP metric is provided in Ref. \cite{Bambhaniya:2021ybs}. Although the equation closely resembles the form presented in Eq. (\ref{hamiltonjacobieq}), a notable distinction lies in the treatment of the function $\mathcal{I} \left(r\,, \theta \right)$. Specifically, Ref. \cite{Bambhaniya:2021ybs} incorporates $\mathcal{I} \left(r\,, \theta \right)$ into the term describing the evolution of the angular coordinate $\theta$ along the geodesic of particles corresponds to the second set of brackets in Eq. (\ref{hamiltonjacobieq}), while excluding it from the term governing the radial coordinate $r$ represented by the first set of brackets in Eq. (\ref{hamiltonjacobieq}). Therefore, it is essential to perform a detailed analysis to determine whether the function $\mathcal{I} \left(r\,, \theta \right)$ predominantly influences the radial coordinate $r$ or the angular coordinate $\theta$. It is equally crucial to ascertain the appropriate association of $\mathcal{I} \left(r\,, \theta \right)$ with either the radial evolution component or the angular evolution component in the Hamilton-Jacobi equation. The dependence of the function $\mathcal{I} \left(r\,, \theta \right)$ on the radial coordinate $r$ and the angular coordinate $\theta$ is illustrated in Fig. \ref{relationrandtheta} for the spin parameter $a = 0.9$ and the deviation parameter $\epsilon_3 = -8$ and $8$.
In Fig. \ref{relationrandtheta}, it is evident that, regardless of whether the deviation parameter $\epsilon_3$ is positive or negative, the variation of the function $\mathcal{I} \left(r\,, \theta \right)$ predominantly depends on the radial coordinate $r$, with minimal influence from the angular coordinate $\theta$. This observation indicates that $\mathcal{I} \left(r\,, \theta \right)$ should be associated with the radial evolution component of the Hamilton-Jacobi equation rather than the angular evolution component. Therefore, $\mathcal{I} \left(r\,, \theta \right)$ should be incorporated into the first bracketed term of the Hamilton-Jacobi equation, as presented in Eq. (\ref{hamiltonjacobieq}). In the subsequent analysis, the Hamilton-Jacobi equation provided in Eq. (\ref{hamiltonjacobieq}) will be utilized to derive the equations of motion for particles in the spacetime described by the JP metric.

Under the limiting condition $\epsilon_3 \to 0$, the JP metric smoothly reduces to the Kerr metric. Since the Carter constant is well-defined in the spacetime described by the Kerr metric, it can be inferred that the Carter constant $\mathcal{K}$ can also be approximately defined within the spacetime described by the JP metric under this limiting condition. With the introduction of this approximately defined Carter constant, the Hamilton-Jacobi equation in Eq. (\ref{hamiltonjacobieq}) can be separated into the following two distinct components as
\bea
    \frac{\left(\Delta+h a^2 \sin ^2 \theta\right)^2}{\Delta(1+h)}\left(\frac{\partial S_r}{\partial r}\right)^2-\frac{\left[\left(r^2+a^2\right) \mathcal{E}-a \mathcal{L}\right]^2}{\Delta}+\left[(\mathcal{L}-a \mathcal{E})^2+m^2 r^2\right] + \mathcal{I} \left(r\,, \theta \right) = -\mathcal{K}
\eea
and
\bea
    \frac{\left(\Delta+h a^2 \sin ^2 \theta\right)}{\Delta}\left(\frac{\partial S_\theta}{\partial \theta}\right)^2+\left(a^2 \cos ^2 \theta+\frac{a^2 \sin ^2 \theta h}{\Delta} \Sigma\right) m^2+\left(\mathcal{L}^2 \csc ^2 \theta-a^2 \mathcal{E}^2\right) \cos ^2 \theta=\mathcal{K}\,.
\eea
Furthermore, the above two equations can be reformulated to express the derivatives of $S_r$ with respect to the radial coordinate $r$ and $S_\theta$ with respect to the angular coordinate $\theta$, respectively. These reformulated expressions are given as
\begin{equation}\label{partialsrr}
    \begin{split}
        \left(\frac{\partial S_r}{\partial r}\right)^2&=\frac{1+h}{\left(\Delta+h a^2 \sin ^2 \theta\right)^2}\mathcal{R}(r)
    \end{split}
\end{equation}
and 
\begin{equation}\label{partialsthetatheta}
    \begin{split}
        \left(\frac{\partial S_\theta}{\partial \theta}\right)^2 = \frac{\Delta}{\left(\Delta+h a^2 \sin ^2 \theta\right)}\Theta(\theta)\,,
    \end{split}
\end{equation}
where $\mathcal{R} (r)$ represents the radial potential of particles, and $\Theta (\theta)$ denotes the angular potential of particles. The explicit forms of these potential functions are
\begin{equation}\label{radialpotential}
    \begin{split}
        \mathcal{R}(r) = \left[\left(r^2+a^2\right) \mathcal{E}-a \mathcal{L}\right]^2-\Delta\left[\mathcal{K}+(\mathcal{L}-a \mathcal{E})^2+ m^2 r^2 \right] + \tilde{\mathcal{I}} (r\,, \theta)
    \end{split}
\end{equation}
and 
\begin{equation}
    \begin{split}
        \Theta(\theta)&=\mathcal{K}-\left(\mathcal{L}^2 \csc ^2 \theta-a^2 \mathcal{E}^2\right) \cos ^2\theta-\left(a^2 \cos ^2 \theta+\frac{a^2 \sin ^2 \theta h}{\Delta} \Sigma\right) m^2\,.
    \end{split}
\end{equation}
The specific expression of the function $\tilde{\mathcal{I}} (r\,, \theta)$ in Eq. (\ref{radialpotential}) is 
\begin{equation}
    \begin{split}
        \tilde{\mathcal{I}} (r\,, \theta) = - \Delta \mathcal{I} (r\,, \theta) = -\frac{\mathcal{E}^2 h \Sigma^2}{(1 + h)}\,.
    \end{split}
\end{equation}
Integrating Eqs. (\ref{partialsrr}) and (\ref{partialsthetatheta}) with respect to the coordinates $r$ and $\theta$, respectively, and substituting the resulting integrated expressions into Eq. (\ref{formalexpaction}), the action $S$ in the Hamiltonian-Jacobi equation can be expressed as
\bea\label{expressionaction}
    S=\frac{1}{2} m^2 \lambda-\mathcal{E} t+\mathcal{L} \phi+\int^r\frac{\sqrt{1+h}}{\Delta+h a^2 \sin^2\theta}\sqrt{\mathcal{R}(r)}dr+\int^\theta\sqrt{\frac{\Delta}{\Delta+h a^2 \sin^2\theta}}\sqrt{\Theta(\theta)}d\theta\,.
\eea
According to the definition of the four momentum of a particle in Eq. (\ref{deffourmomentum}), all components of the four momentum in the spacetime described by the JP metric can be expressed as
\bea\label{allcomponentsmomentum}
    \begin{aligned}
        &p_t=-\mathcal{E}\,,\\
        &p_r=\frac{\partial S}{\partial r}=\frac{\sqrt{1+h}}{\Delta+h a^2 \sin ^2 \theta}\sqrt{\mathcal{R}(r)}\,,\\
        &p_\theta=\frac{\partial S}{\partial \theta}=\sqrt{\frac{\Delta}{\left(\Delta+h a^2 \sin ^2 \theta\right)}}\sqrt{\Theta(\theta)}\,,\\
        &p_\phi=\mathcal{L}\,.
    \end{aligned}
\eea
Using the expression for the action in Eq. (\ref{expressionaction}) and the component expressions for the four-momentum in Eq. (\ref{allcomponentsmomentum}), the geodesic equations in the spacetime described by the JP metric along the four coordinate directions can be written as
\bea
\label{geo}\label{eomcoordirect}
    \begin{aligned}
        &\Sigma \frac{dt}{d\lambda}=\frac{r^2+a^2}{\Delta+a^2 h \sin ^2 \theta}\left[\left(r^2+a^2\right) \mathcal{E}-a \mathcal{L}\right]+\frac{\Delta a}{\Delta+a^2 h \sin ^2 \theta}\left[\mathcal{L}-a \mathcal{E} \sin ^2 \theta\right] - \frac{\mathcal{E}h\Sigma^2}{2 (\Delta+a^2 h \sin ^2 \theta)}\,,\\
        &\Sigma\sqrt{1+h}\frac{dr}{d\lambda}=\sqrt{\mathcal{R}(r)}\,,\\
        &\Sigma \sqrt{\frac{\Delta+a^2 h \sin ^2 \theta}{\Delta}}\frac{d\theta}{d\lambda}=\sqrt{\Theta(\theta)}\,,\\
        &\Sigma\frac{d\phi}{d\lambda}=\frac{a}{\Delta+a^2 h \sin ^2 \theta}\left[\left(r^2+a^2\right) \mathcal{E}-a \mathcal{L}\right]+\frac{\Delta}{\Delta+a^2 h \sin ^2 \theta}\left[\frac{\mathcal{L}}{\sin ^2 \theta}-a \mathcal{E}\right]\,.
    \end{aligned}
\eea
The shadow of the JP black hole can be determined from the photon equations of motion presented in Eq. (\ref{eomcoordirect}) within the corresponding spacetime. The subsequent sections outline the theoretical framework for calculating the JP black hole shadow, emphasizing both the approximate analytical and the numerical approaches.

\subsection{Theoretical foundations for calculating the JP black hole shadow}\label{Analytical Calculation}

Building on the derived equations of motion, we calculate the shadow of the JP black hole as a test of the no-hair theorem. The shadow is formed by photons originating from infinity and the vicinity of the black hole, propagating toward and detected by a distant observer. Photons arriving from infinity are deflected by gravitational lensing and predominantly contribute to the background brightness of the shadow image. Conversely, photons originating near the photon sphere significantly define the structure of the black hole shadow. In the vicinity of the photon sphere, photon trajectories are inherently unstable and can exhibit two modes of motion under perturbations. In the first mode, photons are captured by the black hole, forming the dark region of the shadow as they fail to reach distant observers. In the second mode, photons escape the vicinity of the photon sphere and reach distant observers, delineating the boundary of shadow, commonly known as the critical curve on the image of black hole shadow. The configuration of the critical curve reflects the fundamental feature of black hole shadows, while the geometric properties of the curve encapsulate essential information about the characteristics of black hole shadows. Therefore, investigating the JP black hole shadow focuses on calculating the critical curve and analyzing its geometric configuration to gain insights into the properties of the JP black hole shadow. 

In calculating black hole shadows, it is sufficient to focus solely on the motion of photons within the spacetime. It implies that the parameter $m$ in the derived equations of motion for particles is set to zero, i.e., $m = 0$, ensuring the equations exclusively describe photon trajectories. Under the limiting condition $\epsilon_3 \to 0$, three conserved quantities can be identified along the trajectories of photons in the spacetime described by JP metric, which are the energy $\mathcal{E}$, the angular momentum $\mathcal{L}$, and the Carter constant $\mathcal{K}$. Although the JP metric introduces deviations from the Kerr metric, the unstable spherical photon orbits at $r = \text{constant}$ persist outside JP black holes. These spherical photon orbits collectively form a region known as the photon region. Since the trajectories of these spherical photon orbits are independent of the energy of photons, two dimensionless conserved quantities, $\xi$ and $\eta$, normalized by the energy $\mathcal{E}$, are introduced to characterize the properties of the photon orbits. These conserved quantities are expressed as
\bea
    \xi=\frac{\mathcal{L}}{\mathcal{E}}\,, \qquad \eta=\frac{\mathcal{K}}{\mathcal{E}^2}\,.
\eea
These two conserved quantities are referred to as the impact parameters of photons and are used to characterize the orbits of photons in the spacetime. Two photons share the same trajectory if they have identical values of $\eta$ and $\xi$. In studying the shadow of JP black holes, we primarily focus on photons executing circular motion on the equatorial plane, defined by the angular coordinate $\theta = \pi / 2$. By analysing the properties of roots from the equations of the radial potential $\mathcal{R} (r)$ and the angular potential $\Theta (\theta)$ in the equation of motion of photons, as well as the relationship between these roots and the conserved quantities $\xi$ and $\eta$, the conditions for the radial potential governing photons moving along unstable circular orbits on the equatorial plane of the JP black hole can be expressed as 
\bea\label{randpartialreq0}
    \mathcal{R}(r) = \frac{d \mathcal{R}(r)}{dr}=0
\eea
and
\begin{equation}\label{partial2rrle0}
    \begin{split}
        \frac{d^2 \mathcal{R}(r)}{dr^2} \leq 0\,.
    \end{split}
\end{equation}
The condition in Eq. (\ref{partial2rrle0}) indicates that the circular photon orbits on the equatorial plane of the JP black hole are unstable. Based on Eq. (\ref{randpartialreq0}), the explicit expressions for the two conserved quantities, $\xi$ and $\eta$, which characterize the photons on spherical orbits in the equatorial plane, can be derived as
\bea
\begin{aligned}\label{expxiretar}
\xi(r)&= -\frac{2 a(a-r)(a+r)\left(r^3+\epsilon_3\right)^2 \pm \sqrt{2A (r)}}{2 a^2(-1+r)\left(r^3+\epsilon_3\right)^2}\,,\\
\eta(r)&=\frac{1}{2 a^3(-1+r)^2\left(r^3+\epsilon_3\right)^2} r^2\left( B (r) \pm 2 \sqrt{2A (r)} \right)\,,
\end{aligned}
\eea
where the specific expressions for the two functions $A (r)$ and $B(r)$ are given as
\bea
    \begin{aligned}
        A(r) & =a^2 r^2\left(r^3+\epsilon_3\right)^2\\
        &\left\{2 r^6\left[a^2+(-2+r) r\right]^2+r^3\left[4 a^4+a^2 r(-15+7 r)+r^2(16+r(-17+5 r))\right] \epsilon_3+2\left(a^2-r\right)^2 \epsilon_3^2\right\}\\
        B(r) & =a r\left\{-2 r^7[5+(-4+r) r]+r^4[-20+(17-5 r) r] \epsilon_3-4 r \epsilon_3^2+4 a^2\left(r^3+\epsilon_3\right)^2\right\}\,.
    \end{aligned}
\eea
It is crucial to emphasize that the expressions for $\xi$ and $\eta$ are functions of both the coordinate $r$ and the deviation parameter $\epsilon_3$. As previously discussed, an analytical calculation of the shadow of JP black holes is only feasible under the limiting condition $\epsilon_3 \to 0$ because the Hamilton-Jacobi equation cannot be separated into variables in the spacetime described by the JP metric. Therefore, to facilitate the analytical calculation of the shadow of JP black holes in this limiting case, the expressions for $\xi$ and $\eta$ should be expanded as a series at $ \epsilon_3 = 0$. The first-order approximate expansions of these conserved quantities are given as
\bea\label{expxietaseries}
\begin{aligned}
\xi(r) = & -\frac{a^2 - r ^2+a r \sqrt{\left(a^2-2 r+r ^2\right)^2}}{a(-1+r ) } \pm \frac{a^2-r^2}{4 a r \sqrt{\left(a^2-2 r +r ^2\right)^2}}\epsilon_3+\mo(\epsilon_3)^2\\
\eta(r) = & \frac{r ^3\left[2 a^2-r^3+4 r^2 +5r\pm 2 \sqrt{\left(a^2-2 r+r ^2 \right)^2}\right]}{a^2(-1+r )^2}+\frac{r \left[\mp a^2\pm r^2-r\sqrt{\left(a^2-2 r+r ^2\right) ^2}\right] }{2 a^2(-1+r ) \sqrt{\left(a^2-2 r+r ^2 \right)^2}}\epsilon_3\\
& +\mo(\epsilon_3)^2\,.
\end{aligned}
\eea
The event horizon of JP black holes is determined by the condition $\Delta (r) = 0$, and the inequality $\Delta (r) > 0$ should be satisfied for the spacetime region outside the event horizon. When the deviation parameter $\epsilon_3$ vanishes, i.e., $\epsilon_3 = 0$, the JP metric naturally reduces to the Kerr metric. Therefore, under the simultaneous conditions $\Delta (r) = 0$ and $\epsilon_3 = 0$, the expressions for the two conserved quantities in Eq. (\ref{expxietaseries}) can be further simplified as
\begin{equation}\label{xietaepsilon3eq0}
    \begin{split}
        \xi(r)=\frac{a^2(1+r)+r^2(r-3)}{a(1-r)}\,, \qquad \eta(r)=\frac{r^3\left[4 a^2-r(r-3)^2\right]}{a^2(1-r)^2}\,.
    \end{split}
\end{equation}
The two expressions in Eq. (\ref{xietaepsilon3eq0}) represent the impact parameters of photon trajectories in Kerr spacetime. The radial extent of the photon region outside the event horizon of JP black holes is determined by the roots of the equation $\eta (r) = 0$. Solving this equation yields two real roots, $r_{\text{ph}^-}$ and $r_{\text{ph}^+}$, which represent the inner and outer boundaries of the radial range of unstable circular photon orbits, respectively. This radial range is mathematically expressed as
\begin{equation}
    \begin{split}
        r_{\text{ph}} \in \left[r_{\text{ph}^-}\,, r_{\text{ph}^+} \right]\,.
    \end{split}
\end{equation}
Since the expressions for $r_{\text{ph}^{\pm}}$ are relatively complex and exert minimal influence on subsequent calculations and analyses, the explicit expressions of these quantities are omitted. Under the limiting condition $\epsilon_3 \to 0$, the inner and outer boundaries of the radial range of the photon region simplify to the expression as
\begin{equation}
    \begin{split}
        r_{\text{ph}^{\pm}} = 2 \left[1 + \cos \left(\frac{2}{3} \arccos \left(\pm \left|a\right| \right) \right) \right]\,.
    \end{split}
\end{equation}
These two radial expressions precisely correspond to the inner and outer boundaries of the radial region defined by the circular photon orbits outside the event horizon of Kerr black holes.

\begin{figure}[htbp]
    \centering
    \includegraphics[width=0.5\textwidth]{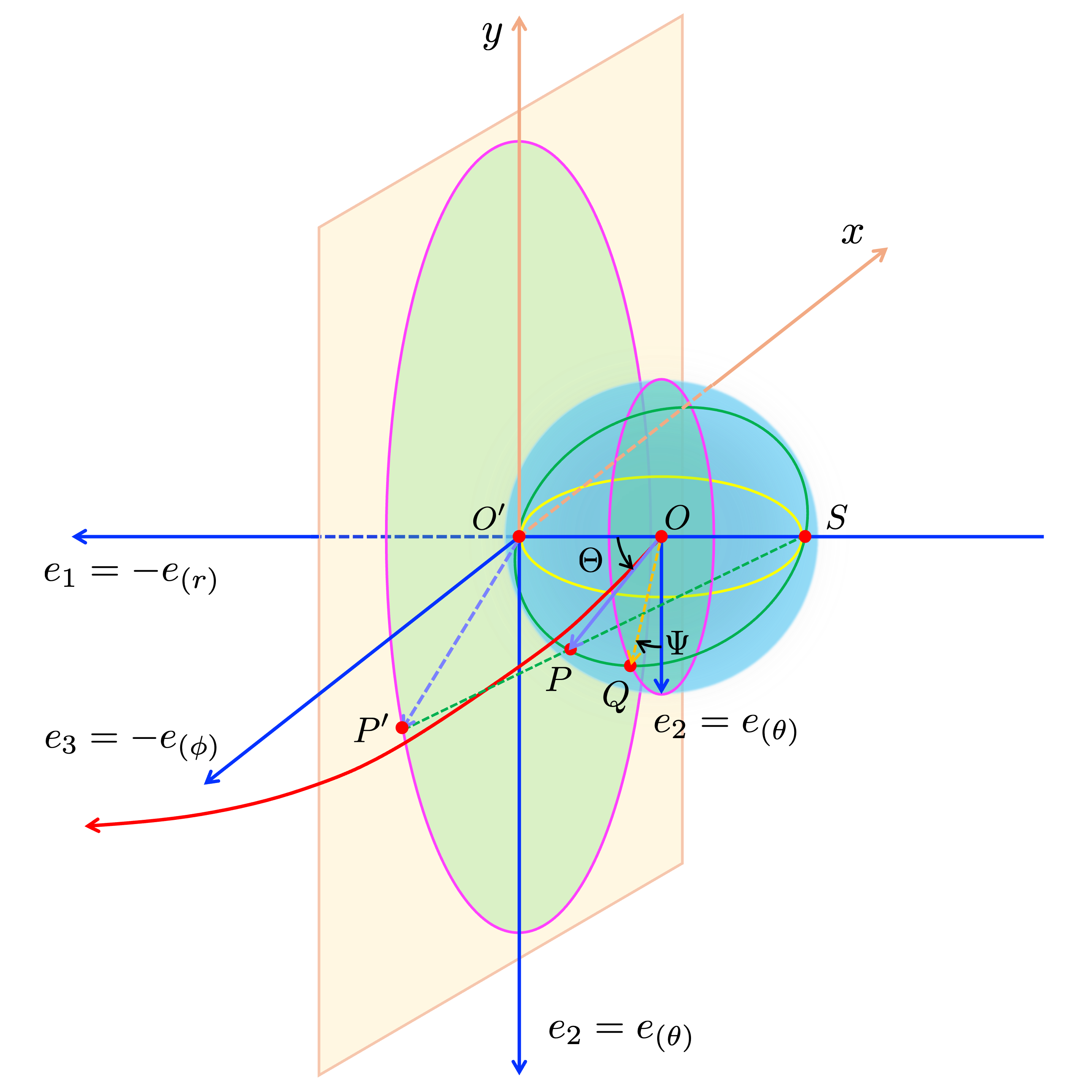}
    \caption{The schematic representation of the ZAMO coordinates is provided. The celestial coordinates $(\Theta\,, \Psi)$ are defined in this framework. The yellow plane denotes the image plane at the location of observer, and the three-momentum of photon can be mapped from the blue sphere onto the image plane using a stereographic projection.}
    \label{zamocoor}
\end{figure}

As mentioned above, photons along null trajectories detected by an observer can be categorized into two types. The first type originates from infinity of spacetime, and the second type emanates from the black hole. These two types of photons are separated by a critical curve on the image plane of observer, which delineates the boundary of the black hole shadow and is commonly referred to as the photon ring. Investigating the black hole shadow involves analyzing the properties of this critical curve on the image plane from the perspective of observer. Therefore, the focus of the subsequent study is to calculate and visualize this critical curve for the JP black hole. To proceed, the position of observer in the spacetime described by the JP metric should first be specified. Due to the symmetry inherent in the spacetime described by the JP metric, the location of the observer is fully determined by the coordinates $(r_0, \theta_0)$, where $\theta_0$ denotes the inclination angle between the observer and the rotational axis of the JP black hole. The coordinates of observer are further refined by assuming that the observer resides in a coordinate system whose basis vectors are defined as
\bea\label{zamocoorbasis}
\hat{e}_{(0)}=\zeta \partial_t+
\gamma \partial_\phi, \quad \hat{e}_{(1)}=-\frac{\partial_r}{\sqrt{g_{r r}}}, \quad \hat{e}_{(2)}=\frac{\partial_\theta}{\sqrt{g_{\theta \theta}}}\,, \quad \hat{e}_{(3)}=-\frac{\partial_\phi}{\sqrt{g_{\phi \phi}}}\,,
\eea
where
\begin{equation}
    \begin{split}
        \zeta=\frac{g_{\phi \phi} }{\sqrt{g_{\phi \phi}\left(g_{\phi t}^2-g_{\phi \phi} g_{t t}\right)}}\,, \qquad \gamma=\frac{-g_{\phi t} }{\sqrt{g_{\phi \phi}\left(g_{\phi t}^2-g_{\phi \phi} g_{t t}\right)}}\,.
    \end{split}
\end{equation}
In the coordinate basis, $\hat{e}_{(0)}$ is the timelike basis vector, and the four-velocity of observer, $\hat{u}$, is aligned with this vector, i.e., $\hat{u} = \hat{e}_{(0)}$. The remaining basis vectors are spacelike, and all basis vectors are normalized and mutually orthogonal in the chosen coordinate system. Furthermore, the observer is assumed to be locally static in this coordinate system. Since the timelike vector $\hat{e}_{(0)}$ is orthogonal to the Killing vector $\partial_\phi$, i.e., $\hat{e}_{(0)} \cdot \partial_\phi = 0$, the observer in this locally static coordinates at spatial infinity of spacetime possesses zero angular momentum. Consequently, this coordinate system is referred to as the zero angular momentum observers (ZAMO) coordinates. A schematic representation of the ZAMO coordinates is shown in Fig. \ref{zamocoor}.
In Fig. \ref{zamocoor}, the observer is positioned at the point $O$. The red line depicts the trajectory of a light ray, with the arrow indicating the direction of photon travel along the path. The vector $\overrightarrow{OP}$ denotes the projection of the tangent vector to the null trajectory at point $O$ onto the three-dimensional spatial plane. This vector is commonly referred to as the three-momentum of the photon. To accurately describe the positions of photons originating from infinity and the vicinity of black holes in the reference frame of observer, it is necessary to introduce celestial coordinates. A sphere is constructed with its center at point $O$ and a radius determined by the vector $\overrightarrow{OP}$. The diameter $O^\prime S$ is aligned parallel to the coordinate axis $\hat{e}_{(1)}$ and constrained to lie within the equatorial plane. In this configuration, the angle between the line $O O^\prime$ and the vector $\overrightarrow{OP}$ is designated as $\Theta$, representing the first coordinate in the celestial coordinate system. The sphere intersects the plane defined by $O O^\prime$ and $OP$ to form a great circle passing through the points $P$, $S$, and $O^\prime$. This great circle intersects another great circle, perpendicular to the basis vector $\hat{e}_{(1)}$, at two points. One of these intersection points, located near $P$, is designated as $Q$. The angle between the vector $\overrightarrow{OQ}$ and the basis vector $\hat{e}_{(2)}$ is defined as $\Psi$, representing the second coordinate in the celestial coordinate system. Therefore, the celestial coordinate system $\left(\Theta\,, \Psi \right)$ is fully established in the ZAMO coordinate system of the observer \cite{Hu:2020usx,Zhong:2021mty}.

According to the JP metric in Eq. (\ref{metric}), the four-momentum of photons in the spacetime described by BL coordinates is expressed as
\begin{equation}
    \begin{split}
        p^\mu = \left(\dot{t}\,, \dot{r}\,, \dot{\theta}\,, \dot{\phi} \right)\,,
    \end{split}
\end{equation}
where the dot denotes the differentiation with respect to the affine parameter $\lambda$, and the symbol $s$ represents the position vector of the photon along this null trajectory. The tangent vector to the null trajectory, expressed as the derivative of $s$ with respect to $\lambda$, is expressed as $\dot{s}$. Consequently, the four-momentum $p^\mu$ is equivalent to $\dot{s}$. The components of the four-momentum $\dot{s}$ in BL coordinates can further be written as
\begin{equation}\label{vec1}
    \begin{split}
        \dot{s}=\dot{t} \partial_t+\dot{r} \partial_r+\dot{\theta} \partial_\theta+\dot{\phi} \partial_\phi\,.
    \end{split}
\end{equation}
Meanwhile, using the celestial coordinates $\left(\Theta\,, \Psi \right)$ defined in the ZAMO coordinate system, the components of the four-momentum of photon can be expressed as
\bea
\label{vec2}
\dot{s}=|\overrightarrow{O P}|\left(- \hat{e}_{(0)}+\cos \Theta \hat{e}_{(1)}+\sin \Theta \cos \Psi \hat{e}_{(2)}+\sin \Theta \sin \Psi \hat{e}_{(3)}\right)\,.
\eea
Using the expressions for the four-momentum of photon in Eqs. (\ref{vec1}) and (\ref{vec2}), along with the definitions of energy $\mathcal{E}$ and angular momentum $\mathcal{L}$ in spacetime, and the normalization condition of the ZAMO coordinate basis $\hat{e}_{(0)} \cdot \hat{e}_{(0)} = -1$, the scale factor $|\overrightarrow{O P}|$ in Eq. (\ref{vec2}) is given by
\bea
 \quad |\overrightarrow{O P}| = \gamma \mathcal{L}-\zeta \mathcal{E}\,.
\eea
By substituting the expression from Eq. (\ref{zamocoorbasis}) into Eq. (\ref{vec2}) and comparing the coefficients of $\partial_r$ and $\partial_\phi$ in Eqs. (\ref{vec1}) and (\ref{vec2}), the specific expressions for $\dot{r}$ and $\dot{\phi}$ are obtained as
\bea
\begin{aligned}
    \dot{r} = & |\overrightarrow{O P}| \cos \Theta \frac{-1}{\sqrt{g_{r r}}}\,,\\
    \dot{\phi} = & |\overrightarrow{O P}| \left(-\gamma+\sin \Theta \sin \Psi\frac{-1}{\sqrt{g_{\phi \phi}}}\right)\,.
\end{aligned}    
\eea
Therefore, the celestial coordinates can be explicitly represented using trigonometric functions as
\bea
    \begin{aligned}\label{expthetaphi}
        \cos \Theta &=\frac{\sqrt{g_{r r}} \dot{r}}{\zeta \mathcal{E}-\gamma \mathcal{L}}\,,\\
        \sin \Psi &=\frac{\sqrt{g_{\phi \phi}}}{\sin\Theta}\left(\frac{\dot{\phi}}{\zeta \mathcal{E}-\gamma \mathcal{L}}-\gamma\right)\,.
    \end{aligned}
\eea
The specific expressions for $\dot{r}$ and $\dot{\phi}$ can be derived from the component form of the equations of motion expressed in the BL coordinates in Eq.\eqref{geo}. Building on this foundation, the approach of stereographic projection \cite{Grenzebach:2014fha, Hu:2020usx}, also referred to as the fisheye camera model, is employed to visualize the image of the JP black hole shadow. This method has been used to determind the black hole shadow and images in various works \cite{Liu:2024soc,He:2024amh,Zhang:2022osx,Li:2024ctu,Hou:2022eev,Zhang:2024lsf}.  As depicted in Fig. \ref{zamocoor}, a plane is positioned adjacent to the left side of the sphere, with Cartesian coordinates $O^\prime x y$ established on the plane. The optical center of the camera is located at point $S$, and the point $P$ on the sphere is mapped onto the plane as the image point $P^\prime$ using the stereographic projection method. Based on the geometric relationships depicted in Fig. \ref{zamocoor}, the coordinates of the image point $P^\prime$ on the plane can be represented in the Cartesian coordinate system can be expressed as
\bea
\begin{aligned}\label{stereographicproject}
& x_{p^\prime} = -2 |\overrightarrow{O P}| \tan \left(\frac{\Theta }{2}\right) \sin \Psi\,, \\
& y_{p^\prime} = -2 |\overrightarrow{O P}| \tan \left(\frac{\Theta}{2}\right) \cos \Psi\,.
\end{aligned}
\eea
The critical curve encapsulates the essential structural features of a black hole shadow, making its calculation and visualization sufficient for a thorough analysis of the properties of shadow. This curve is determined by photons originating from the photon sphere orbits outside the event horizon, propagating to a distant observer, and being recorded as observable data. The trajectories of these photons are determined by the impact parameters $\xi (r)$ and $\eta (r)$, evaluated at the photon sphere radius $r_{\text{ph}}$. In the reference frame of observer, the three-momentum of photons near the photon sphere is expressed in celestial coordinates within the ZAMO framework and further projected onto the two-dimensional image plane $O^{\prime}xy$ in Fig. \ref{zamocoor}. This projection allows the construction of the critical curve of the JP black hole shadow as perceived by the distant observer. To construct the critical curve of the JP black hole shadow, the explicit expressions for the impact parameters $\xi (r)$ and $\eta (r)$ at the photon sphere radius $r_{\text{ph}}$ are derived using Eq. (\ref{expxiretar}). Subsequently, the relationship between the celestial coordinates and the impact parameters $\xi (r_{\text{ph}})$ and $\eta (r_{\text{ph}})$ is established through Eq. (\ref{expthetaphi}). Finally, the three-momentum of photons is projected onto the image plane the stereographic projection in Eq. (\ref{stereographicproject}). This systematic process outlines an analytical framework for calculating the critical curve of the JP black hole shadow, assuming that the Hamilton-Jacobi equation governing photon motion permits variable separation. However, as previously discussed, the Hamilton-Jacobi equation in the spacetime described by the JP metric does not permit exact variable separation. Therefore, it requires the application of the numerical computation technique to precisely determine the shadow of the JP black hole.

\begin{figure}[htbp]
    \centering
    \begin{minipage}[b]{0.45\textwidth}
        \centering
        \includegraphics[width=\textwidth]{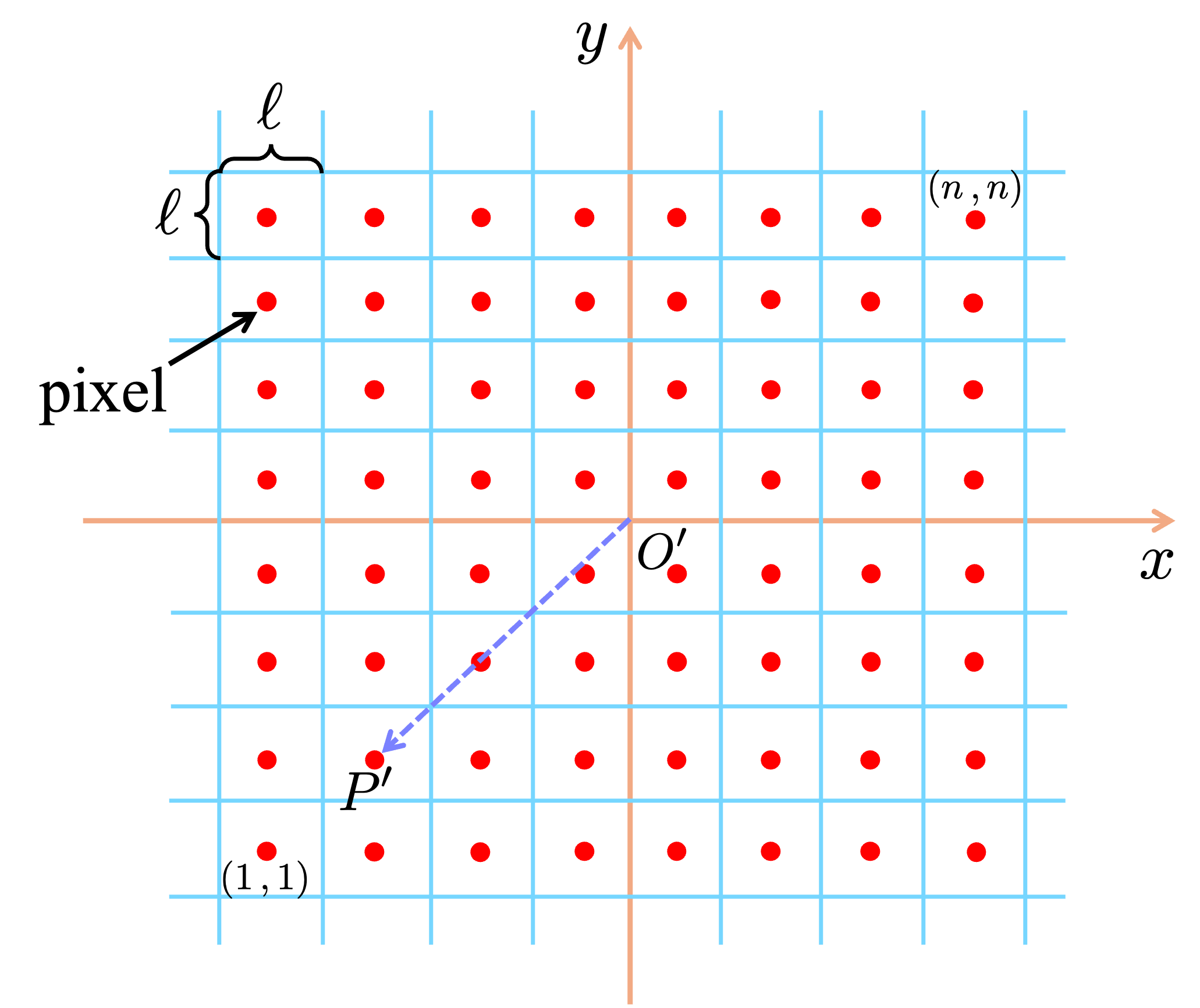}
        \caption*{(a)} 
    \end{minipage}
    \begin{minipage}[b]{0.38\textwidth}
        \centering
        \includegraphics[width=\textwidth]{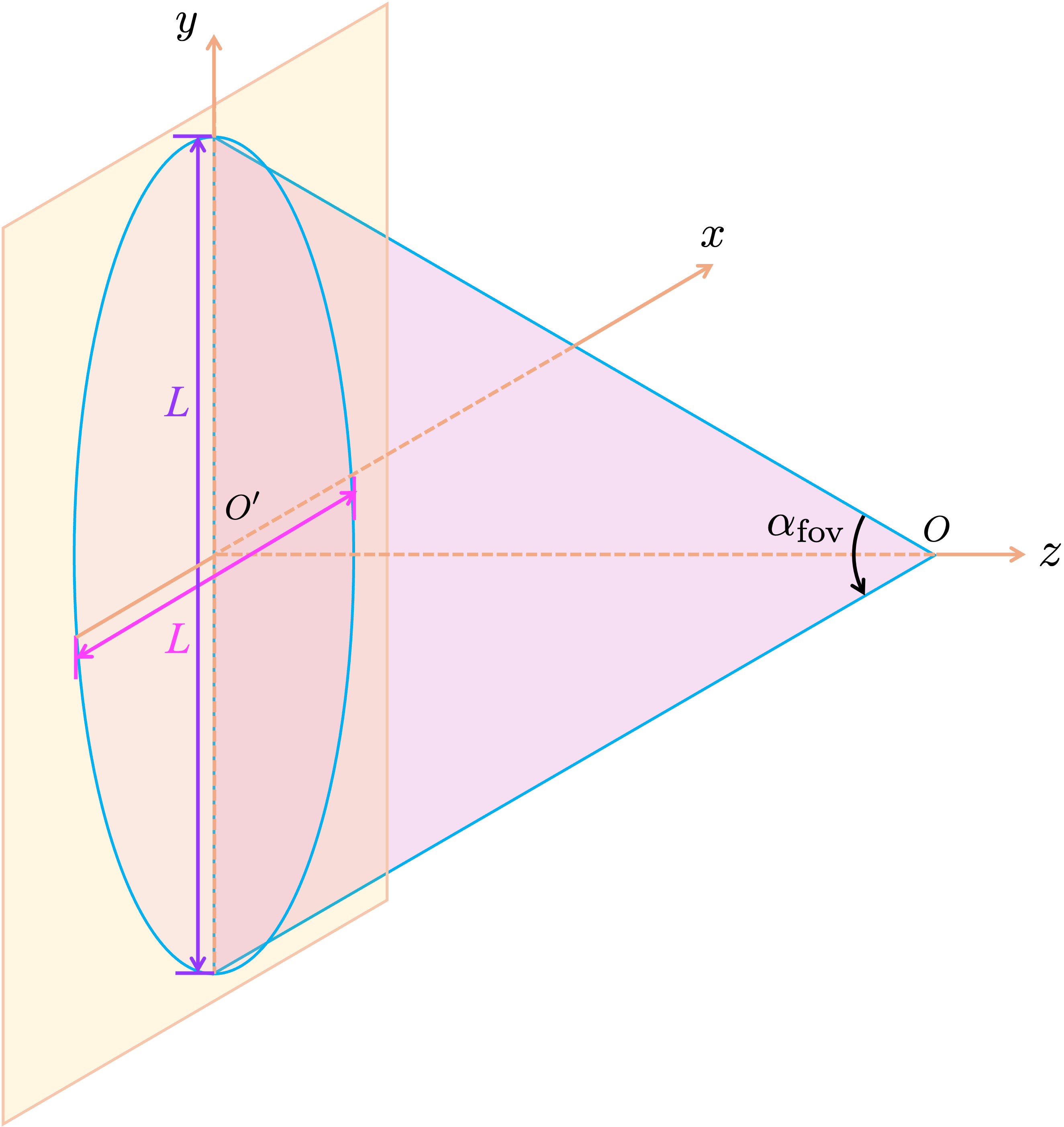}
        \caption*{(b)}
    \end{minipage}
    \caption{(a) The discretized grid points on the image plane are defined in the reference frame of observer. (b) The maximum angular extent $\alpha_{\text{fov}}$ defines the angular range corresponding to the field of view captured by the camera.}
    \label{gridandfov}
\end{figure}

The numerical computation of the JP black hole shadow is primarily performed using the backward ray-tracing method. The process begins by projecting the three-momentum of photons, originating either from infinity or the vicinity of the black hole and reaching the observer, onto the image plane in the coordinates of observer, as described by Eq. (\ref{expxietaseries}). Unlike the analytical approach, the numerical method requires discretizing the image plane into a grid, where each grid point corresponds to a pixel in the resulting black hole shadow image. A schematic representation of the discretized image plane is shown in Fig. \ref{gridandfov} (a).
Additionally, determining the field of view of camera is essential for capturing the black hole shadow. The field of view corresponds to the maximum angular extent the camera can capture, defined by angles in the $xO^\prime z$ and $yO^\prime z$ planes, as shown in Fig. \ref{gridandfov} (b). For simplicity, a square region with a side length $L$, aligned along the $x$- and $y$-directions, is selected on the image plane. The relationship between the side length $L$ of the square region and the maximum angular extent of camera $\alpha_{\text{fov}}$ is schematically illustrated and can be expressed as
\begin{equation}
    \begin{split}
        L = 2 \left|\overrightarrow{OP} \right| \tan \left(\frac{\alpha_{\text{fov}}}{2} \right)\,.
    \end{split}
\end{equation}
When the image plane is discretized into an $n \times n$ grid, the side length of each pixel is given by
\begin{equation}
	\begin{split}
		\ell = \frac{2 \left|\overrightarrow{O P} \right|}{n} \tan \frac{\alpha_{\text{fov}}}{2}\,.
	\end{split}
\end{equation}
The pixels on the image plane are indexed by $(i\,, j)$, with the bottom-left corner designated $(1\,, 1)$ and the top-right corner as $(n\,, n)$. It indicates that the values of indices $i$ and $j$ range from $1$ to $n$. The Cartesian coordinates of any arbitrary pixel $P^\prime$ on the image plane are expressed as
\begin{equation}\label{pixelandcartesian}
	\begin{split}
		x_{p^\prime} = \ell \left(i - \frac{n + 1}{2} \right)\,, \qquad y_{p^\prime} = \ell \left(j - \frac{n + 1}{2} \right)\,.
	\end{split}
\end{equation}
Using Eqs. (\ref{stereographicproject}) and (\ref{pixelandcartesian}), the relationship between the celestial coordinates and the pixel indices on the image plane can be mathematically expressed as
\begin{equation}\label{relationthetaphiij}
	\begin{split}
		\tan \Psi = & \frac{i - \frac{n + 1}{2}}{j - \frac{n + 1}{2}}\,,\\
        \tan \frac{\Theta}{2} = & \frac{\tan \frac{\alpha}{2}}{n} \left[\left(i - \frac{n + 1}{2} \right)^2 + \left(j - \frac{n + 1}{2} \right)^2 \right]^{\frac{1}{2}}\,.
	\end{split}
\end{equation}
When numerically computing the black hole shadow via the backward ray-tracing method, it is essential to establish the grid resolution of the image plane, $n$, and the field of view, $\alpha_{fov}$. The indices of each pixel $(i, j)$ on the image plane are converted into celestial coordinates $(\Theta, \Phi)$ using Eq. (\ref{relationthetaphiij}). Subsequently, the corresponding four-momentum of the photon is transformed from the ZAMO coordinates to the BL coordinates through Eqs. (\ref{vec1}) and (\ref{vec2}). These transformations establish the initial conditions, which include the pixel position and the four-momentum of the photon, necessary for computing the trajectory of the photon through the backward ray-tracing method. The photon trajectory is computed by solving the Hamiltonian canonical equations,
\begin{equation}
    \begin{split}
       \dot{x}^\mu = \frac{\partial H}{\partial p_\mu}\,, \qquad \dot{p}_\mu = - \frac{\partial H}{\partial x^\mu}\,,
    \end{split}
\end{equation}
to ensure computational accuracy, as the Hamilton-Jacobi equation does not allow strict separability in an arbitrary spacetime, where $x^{\mu} = \left(t\,, r\,, \theta\,, \phi \right)$. By performing this procedure to every pixel on the image plane, the black hole shadow can be accurately reconstructed on the image plane. In the subsequent analysis, the shadow of the JP black hole will be computed using both the approximate analytical approach and the numerical method outlined in this section.

\subsection{Approximate analytical and numerical calculation of JP black holes shadow}\label{Numerical Calculation of the Image and Shadow}

The inclusion of the parameter $\epsilon_3$ introduces measurable deviations between the JP and Kerr metrics, which become more pronounced as $\epsilon_3$ increases. These deviations cause photon trajectories in spacetime described by JP metric to diverge significantly from those in spacetime described by the Kerr metric, leading to notable changes in the structure of the JP black hole shadow compared to the Kerr black hole shadow. Additionally, the parameter $\epsilon_3$ alters the Petrov classification of the JP metric from the Petrov type-D, as in the Kerr metric, to the Petrov type-I. The alteration of the metric type in the Petrov classification prevents the definition of the Carter constant in the spacetime described by the JP metric. Consequently, the Hamilton-Jacobi equation governing the motion of photons in the JP spacetime cannot be rigorously solved through variable separation. However, an approximate separation of variables can be achieved by analyzing the correlation between the terms in the Hamilton-Jacobi equation and the coordinates $r$ and $\theta$, as discussed in Sec. \ref{Geodesic Equations}. Based on the approximately separated Hamilton-Jacobi equation, an analytical solution for the shadow of the JP black hole can be derived. Furthermore, the shadow can be accurately reconstructed through the numerical calculation using the backward ray-tracing method \cite{Hu:2020usx}. In the subsequent analysis, we explore the relationship between the shadow of the JP black hole, determined using the approximate variable separation method, and that of the Kerr black hole. Furthermore, we compare the JP black hole shadow obtained through the approximate analytical approach with results from the numerical computation to determine the permissible range of $\epsilon_3$ where the approximate analytical approach remains consistent with the numerical computation. Finally, we systematically examine the effects of varying $\epsilon_3$ on the characteristics of the shadow of JP black hole.

Before analyzing the shadow of the JP black hole, the position of observer in the spacetime described by the JP metric should be specified. For consistency, the observer is fixed at $\left(r_0\,, \theta_0 \right) = \left(300\,, 90^\circ \right)$ throughout the analysis. The critical curve of the JP black hole shadow is defined by the impact parameters of photons, $\xi (r)$ and $\eta (r)$, which are determined by the radial potential $\mathcal{R} (r)$ in Eq. (\ref{radialpotential}) derived from the Hamilton-Jacobi equation for photons. The function $\tilde{\mathcal{I}} (r\,, \theta)$ in the radial potential introduces deviations from the radial potential of photons in Kerr spacetime due to the parameter $\epsilon_3$, which quantifies the difference between the JP black hole and the Kerr black hole. These deviations alter the photon impact parameters and modify the configuration of the critical curve of the JP black hole shadow relative to the Kerr black hole shadow. However, when $\tilde{\mathcal{I}} (r\,, \theta)$ is excluded from $\mathcal{R} (r)$, the radial potential simplifies to the form in the Kerr spacetime, and the critical curve of the black hole shadow coincides exactly with that of a Kerr black hole. To illustrate these effects, we will plot the critical curves of the JP black hole shadow for two cases: one with the function $\tilde{\mathcal{I}} (r\,, \theta)$ included in the radial potential and one without. These curves will be compared with the critical curve of Kerr black hole shadow for intuitive understanding. Furthermore, the exact critical curve of the JP black hole shadow will be calculated using the numerical method, and the results will be systematically compared with those derived from the approximate analytical approach to evaluate the consistency and accuracy of the two methods.

\begin{figure}[htbp]
    \centering
    \begin{minipage}[b]{0.45\textwidth}
        \centering
        \includegraphics[width=\textwidth]{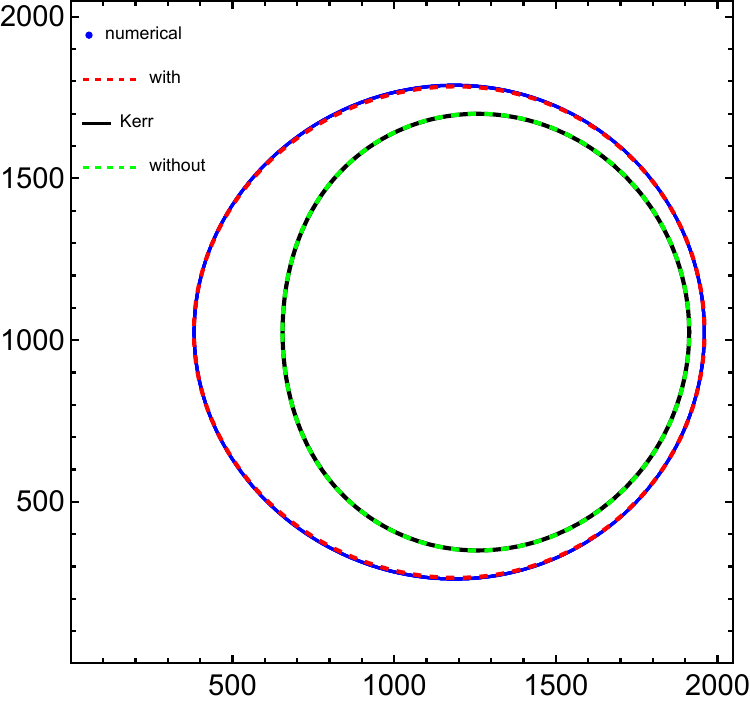}
        \caption*{(a) $a = 0.9\,, \epsilon_3 = - 8$} 
    \end{minipage}
    \begin{minipage}[b]{0.45\textwidth}
        \centering
        \includegraphics[width=\textwidth]{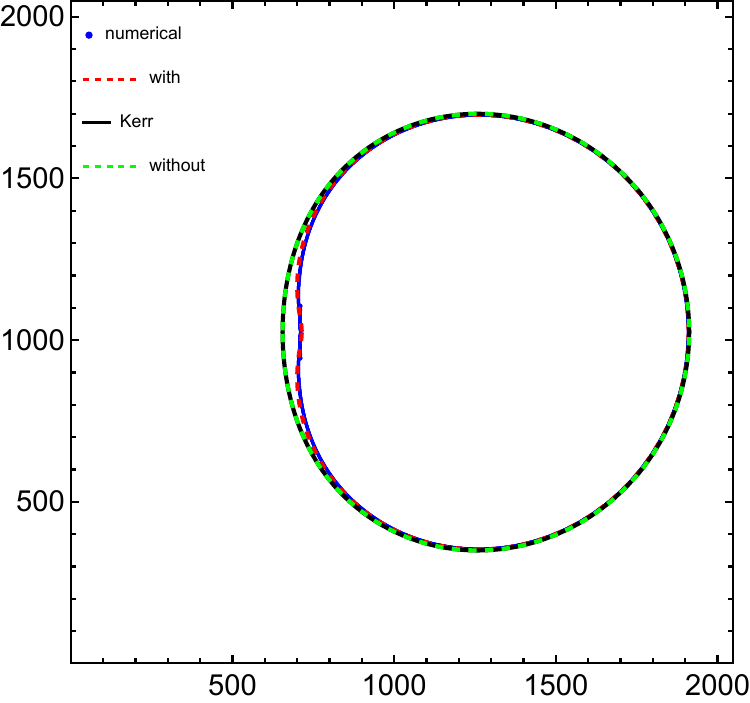}
        \caption*{(b) $a = 0.9\,, \epsilon_3 = 0.3$}
    \end{minipage}
    \caption{The comparative analysis of the critical curves of the JP black hole shadow with a closed event horizon and the Kerr black hole shadow is illustrated. The black solid line represents the critical curve of the Kerr black hole shadow. The green and red dashed lines correspond to the critical curves of the JP black hole shadow derived using approximate analytical methods, with the radial potential $\mathcal{R}(r)$ excluding and including the function $\tilde{\mathcal{I}}(r, \theta)$, respectively. The blue solid line denotes the critical curve of the JP black hole shadow obtained through the numerical computation. Panel (a) depicts the case for the JP black hole with $a = 0.9$ and $\epsilon_3 = -8$, while Panel (b) shows the case for the JP black hole with $a = 0.9$ and $\epsilon_3 = 0.3$.}
    \label{withwithoutikerr}
\end{figure}

The comparisons between the critical curves of the JP black hole shadow with a closed event horizon and those of the Kerr black hole are shown in Fig. \ref{withwithoutikerr}.
In this figure, the black solid curve represents the critical curve of the Kerr black hole shadow, while the dashed lines depict the critical curves of the JP black hole shadow derived using the approximate analytical method. Specifically, the red dashed line corresponds to the case where the radial potential includes the function $\tilde{\mathcal{I}} (r\,, \theta)$, and the green dashed line represents the case where this function is excluded. The blue solid line illustrates the critical curve of the JP black hole shadow obtained through the numerical computation. For $\epsilon_3 < 0$, as shown in Fig. \ref{withwithoutikerr} (a), the critical curve of the JP black hole shadow, calculated with the radial potential excluding the function $\tilde{\mathcal{I}} (r\,, \theta)$, coincides exactly with the Kerr black hole shadow. However, when the function $\tilde{\mathcal{I}} (r\,, \theta)$ is incorporated in the radial potential, the critical curve of the JP black hole shadow expands relative to that of the Kerr black hole shadow, adopting an elliptical profile with its major axis aligned along the $x$-axis. Consequently, the critical curve of the JP black hole derived from the radial potential containing $\tilde{\mathcal{I}} (r\,, \theta)$ exhibits a significant deviation from the critical curve of the Kerr black hole. Furthermore, the shadow of the JP black hole is rigorously calculated using numerical methods, and the corresponding critical curve is reconstructed. These results indicate that when the function $\tilde{\mathcal{I}} (r\,, \theta)$ is incorporated into the expression for the photon radial potential, the JP black hole shadow derived using the approximate analytical approach is entirely consistent with the results obtained through the method of numerical computation. This consistency establishes the reliability of the approximate analytical method as an alternative to numerical approaches for analyzing the JP black hole shadow.

For the case of $\epsilon_3 > 0$ and the event horizon remaining closed, as illustrated in Fig. \ref{withwithoutikerr} (b), the results are analogous to those observed for $\epsilon_3 < 0$. The approximate analytical method is initially employed to compute the shadow of a JP black hole. The results indicate that the critical curve of the JP black hole shadow matches that of the Kerr black hole when the radial potential of photons excludes the function $\tilde{\mathcal{I}} (r\,, \theta)$. However, when the radial potential includes the function, the critical curve of the JP black hole shadow closely resembles that of the Kerr black hole shadow across most regions, except in the vicinity of $y = 200$, where the critical curve of the JP black hole shadow shows a noticeable inward deviation compared to that of the Kerr black hole. Furthermore, the critical curve of the JP black hole shadow obtained through numerical computation aligns precisely with that derived from the approximate analytical approach when the radial potential includes the function $\tilde{\mathcal{I}} (r\,, \theta)$. This agreement demonstrates that, for $\epsilon_3 > 0$ and a closed event horizon, the approximate analytical method reliably substitutes the numerical computation for investigating the JP black hole shadow. In conclusion, when the event horizon of the JP black hole remains closed and the radial potential of photons incorporates the function $\tilde{\mathcal{I}} (r\,, \theta)$, the critical curve of the JP black hole shadow calculated through the approximate analytical method is precisely consistent with the critical curve derived via the numerical computation. This consistency establishes the approximate analytical method as a dependable alternative to numerical approaches for investigating the shadow properties of the JP black hole.

\begin{figure}[htbp]
    \centering
    \begin{minipage}[b]{0.45\textwidth}
        \centering
        \includegraphics[width=\textwidth]{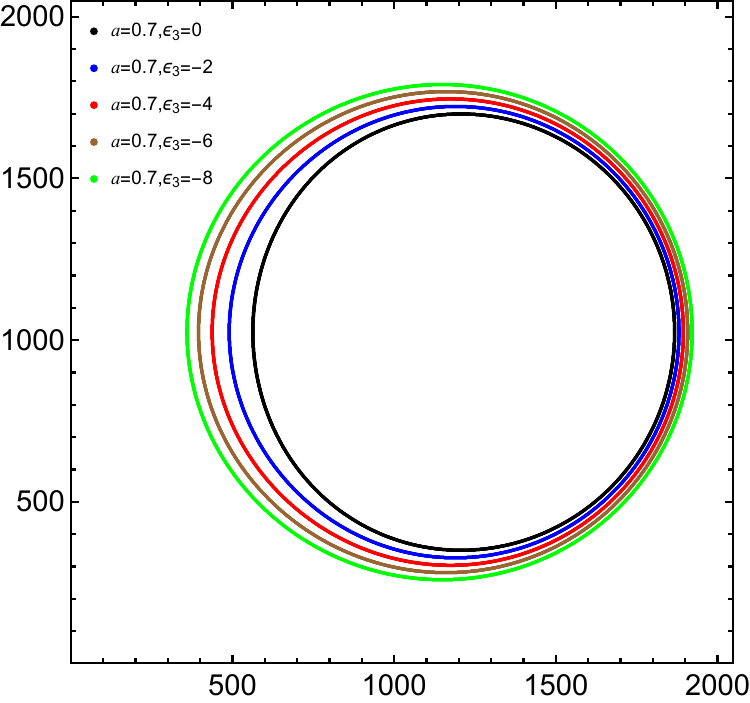}
        \caption*{(a)} 
    \end{minipage}
    \begin{minipage}[b]{0.45\textwidth}
        \centering
        \includegraphics[width=\textwidth]{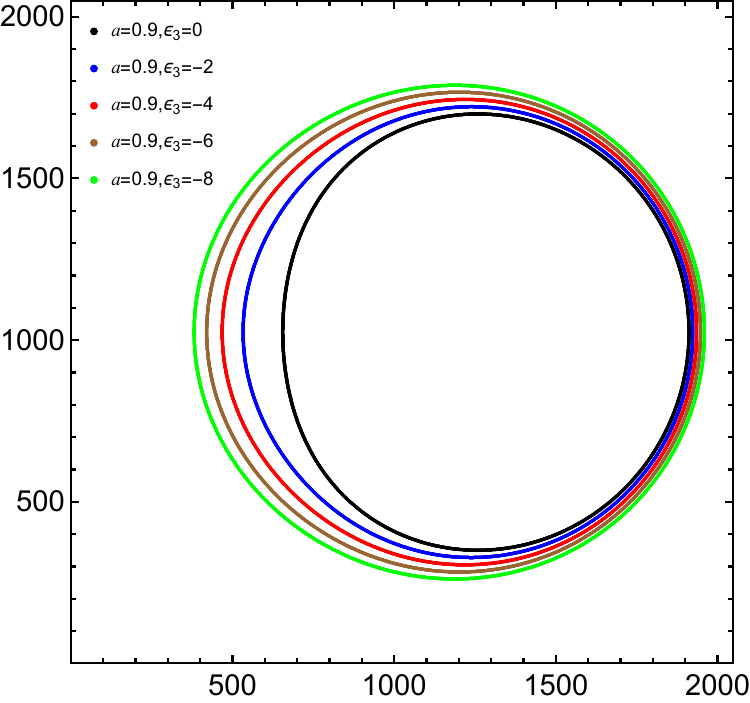}
        \caption*{(b)}
    \end{minipage}
    \caption{The evolution of the critical curve of the JP black hole shadow as the absolute value of the parameter $\epsilon_3$ increases for the negative values of $\epsilon_3$ is illustrated. Panel (a) presents the case for the JP black hole with $a = 0.7$, while Panel (b) describes the case of the JP black hole with $a = 0.9$.}
    \label{jpshadowwithnegativeepsilon}
\end{figure}

\begin{figure}[htbp]
    \centering
    \begin{minipage}[b]{0.45\textwidth}
        \centering
        \includegraphics[width=\textwidth]{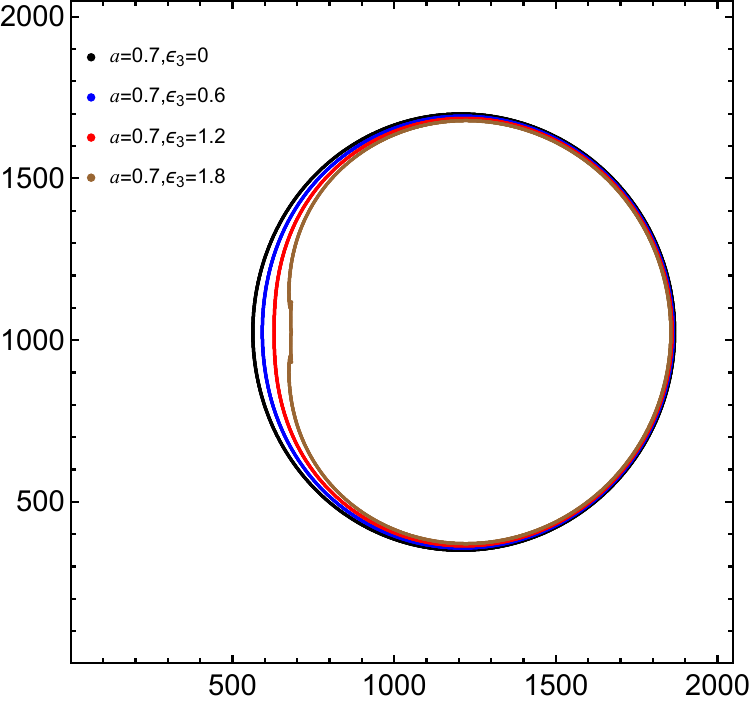}
        \caption*{(a)} 
    \end{minipage}
    \begin{minipage}[b]{0.45\textwidth}
        \centering
        \includegraphics[width=\textwidth]{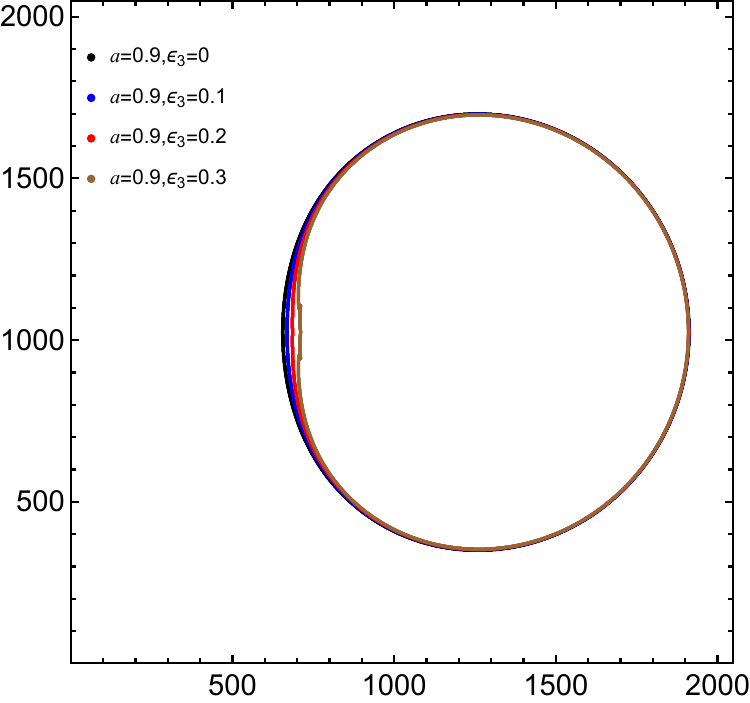}
        \caption*{(b)}
    \end{minipage}
    \caption{The evolution of the critical curve of the JP black hole shadow as the parameter $\epsilon_3$ increases for positive values of $\epsilon_3$, with the event horizon remaining closed, is illustrated. Panel (a) shows the case for the JP black hole with $a = 0.7$, while Panel (b) depicts the case of the JP black hole with $a = 0.9$.}
    \label{jpshadowwithpositiveepsilon}
\end{figure}

Subsequently, we analyze the evolution of the JP black hole shadow with respect to variations in the parameter $\epsilon_3$, under the condition that the event horizon of the JP black hole is remaining closed. For negative values of the parameter $\epsilon_3$, Fig. \ref{jpshadowwithnegativeepsilon} illustrates the variation in the critical curve of the JP black hole shadow as a function of $\epsilon_3$, with the spin parameters of the JP black hole set to $a = 0.7$ and $a = 0.9$.
The results in Fig. \ref{jpshadowwithnegativeepsilon} reveal that for both spin parameters, $a = 0.7$ and $a = 0.9$, the critical curve of JP black hole shadow exhibits a consistent trend of variation as the absolute value of $\epsilon_3$ increases. Specifically, as the absolute value of $\epsilon_3$ increases, the critical curve of the JP black hole shadow progressively expands relative to the Kerr black hole shadow. The contour of the critical curve initially approaches a circle shape before transitioning into an ellipse form with the major axis aligned along the $x$-axis. Conversely, for positive values of the parameter $\epsilon_3$ with the event horizon of the JP black hole remaining closed, Fig. \ref{jpshadowwithpositiveepsilon} illustrates the evolution of the critical curve of the JP black hole shadow as $\epsilon_3$ increases.
The results presented in Fig. \ref{jpshadowwithpositiveepsilon} demonstrate that for the JP black hole with a spin parameter of $a = 0.7$, the critical curve of the JP black hole shadow increasingly exhibits an inward concave deformation relative to the critical curve of the Kerr black hole shadow as the parameter $\epsilon_3$ increases. A similar pattern is observed for $a = 0.9$, with the critical curve displaying inward concavity as $\epsilon_3$ increases. In both cases, the critical curve recedes inward relative to the Kerr black hole shadow as the value of $\epsilon_3$ increases. However, for the case of $a = 0.9$, the inward concave deformation of the critical curve of the JP black hole shadow becomes less pronounced as $\epsilon_3$ increases, resulting in a smaller overall deviation from the Kerr black hole shadow compared to the case of $a = 0.7$. As $\epsilon_3$ increases, the critical curve of the JP black hole shadow for $a = 0.9$ exhibits minimal deviation, rendering it almost indistinguishable from the Kerr black hole shadow. These observations suggest that the magnitude of the spin parameter $a$ mitigates the extent of the deviation introduced by the parameter $\epsilon_3$.

\begin{figure}[htbp]
    \centering
    \begin{minipage}[b]{0.3\textwidth}
        \centering
        \includegraphics[width=\textwidth]{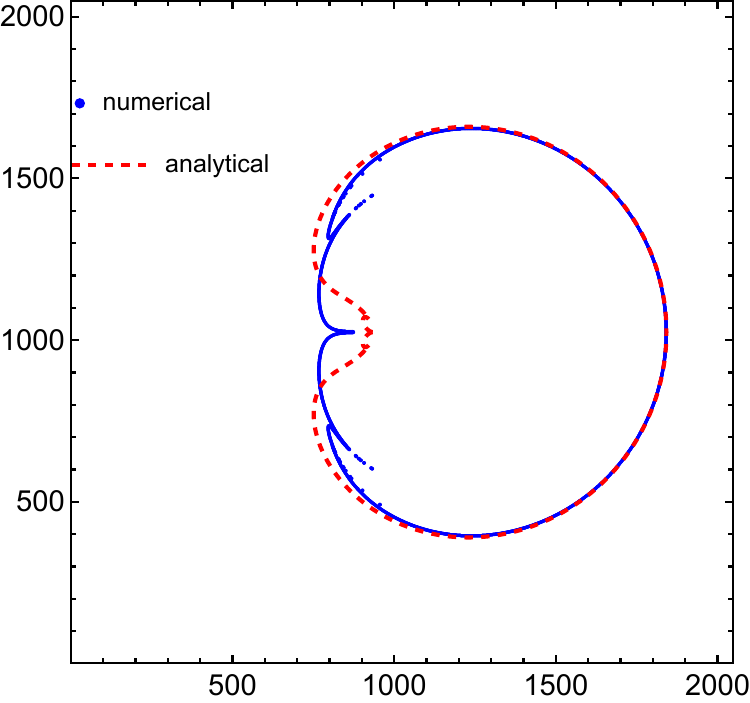}
        \caption*{(a) $a = 0.7\,, \epsilon_3 = 4$}
    \end{minipage}
    \begin{minipage}[b]{0.3\textwidth}
        \centering
        \includegraphics[width=\textwidth]{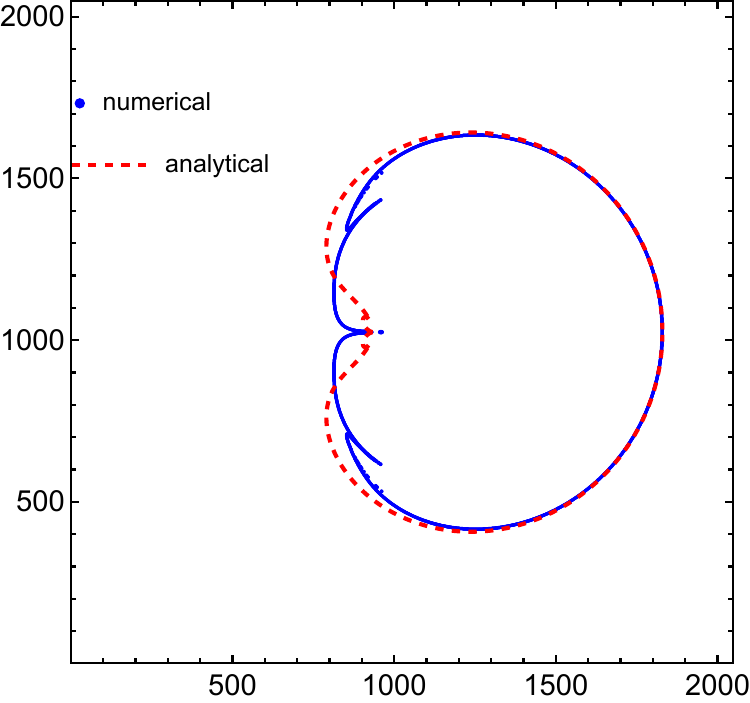}
        \caption*{(b) $a = 0.7\,, \epsilon_3 = 6$}
    \end{minipage}
    \begin{minipage}[b]{0.3\textwidth}
        \centering
        \includegraphics[width=\textwidth]{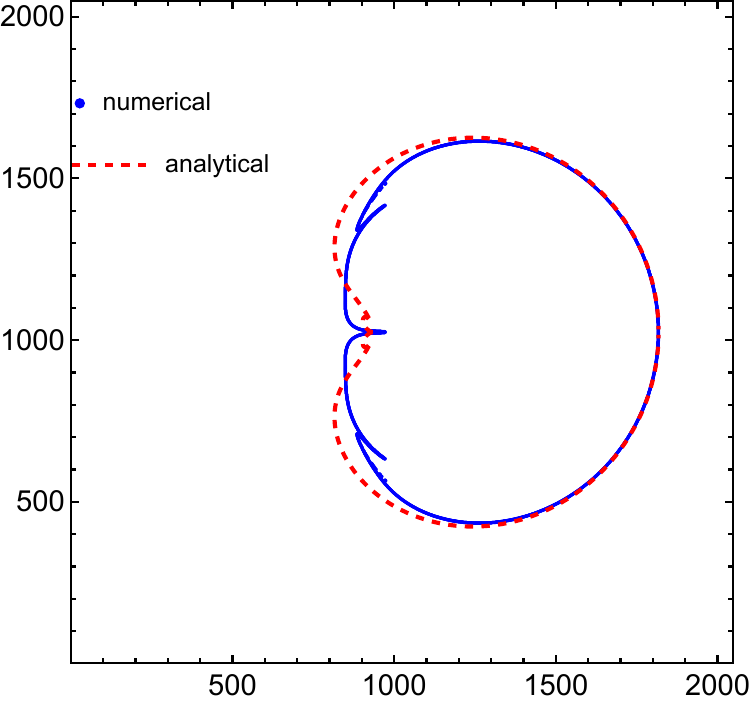}
        \caption*{(c) $a = 0.7\,, \epsilon_3 = 8$}
    \end{minipage}

    \vspace{1em}

    \begin{minipage}[b]{0.3\textwidth}
        \centering
        \includegraphics[width=\textwidth]{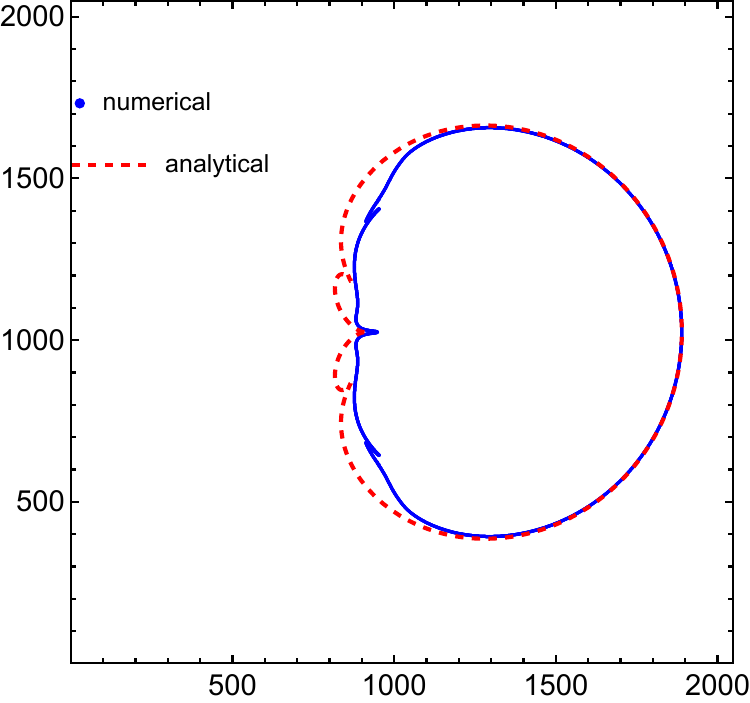}
        \caption*{(d) $a = 0.9\,, \epsilon_3 = 4$}
    \end{minipage}
    \begin{minipage}[b]{0.3\textwidth}
        \centering
        \includegraphics[width=\textwidth]{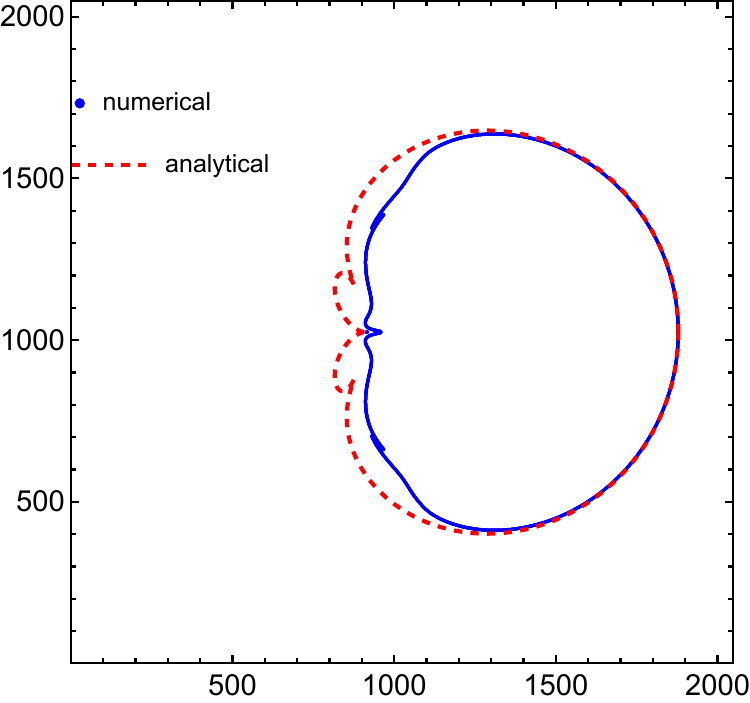}
        \caption*{(e) $a = 0.9\,, \epsilon_3 = 6$}
    \end{minipage}
    \begin{minipage}[b]{0.3\textwidth}
        \centering
        \includegraphics[width=\textwidth]{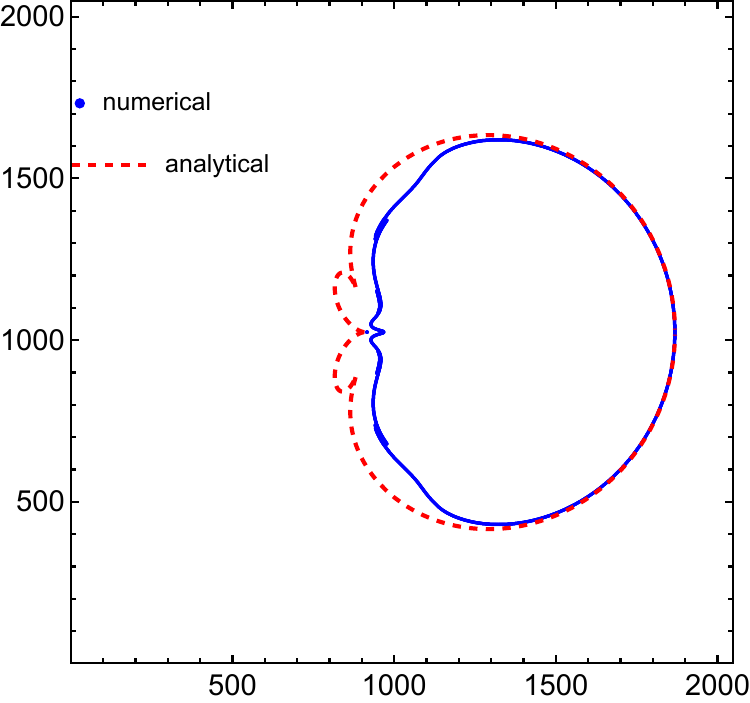}
        \caption*{(f) $a = 0.9\,, \epsilon_3 = 8$}
    \end{minipage}
    \caption{The comparison between the critical curves of the JP black hole shadow derived using the approximate analytical method and those obtained through the numerical computation is illustrated. The red dashed lines represent the critical curves obtained via the approximate analytical method, while the blue solid lines correspond to the results obtained from numerical computations. Panels (a), (b), and (c) show the JP black hole with $a = 0.7$ and $\epsilon_3 = 4$, $6$, and $8$, respectively. Panels (d), (e), and (f) depict the JP black hole with $a = 0.9$ and $\epsilon_3 = 4$, $6$, and $8$, respectively.}
    \label{relationanalynumernonclosed}
\end{figure}

Next, we investigate the properties of the JP black hole shadow under the condition of a non-closed event horizon, where the parameter $\epsilon_3$ consistently assumes positive values. Similar to the case of the JP black hole with a closed event horizon, we first analyze the relationship between the JP black hole shadow calculated using the approximate analytical approach and that obtained through the numerical computation. Figure \ref{relationanalynumernonclosed} presents the critical curves of the JP black hole shadow for spin parameters $a = 0.7$ and $a = 0.9$, with $\epsilon_3$ set to $4$, $6$, and $8$, respectively. These curves are obtained using both the approximate analytical approach and the numerical computation.
Meanwhile, the comparison between the critical curves of the JP black hole shadow obtained through approximate analytical calculations and those derived from the numerical computation is also presented in Fig. \ref{relationanalynumernonclosed}. The red dashed lines represent the critical curves derived using the approximate analytical method, while the blue solid lines correspond to the results from the numerical calculation. The results indicate that, when the event horizon of the JP black hole is not closed, significant discrepancies persist between the critical curves obtained by these two methods, irrespective of the value of $\epsilon_3$. This inconsistency demonstrates that the approximate analytical method fails to accurately capture the characteristics of the JP black hole shadow and cannot serve as a substitute for the numerical computation when the event horizon is non-closed. Consequently, the approximate analytical approach is entirely unsuitable for studying JP black hole shadows under such conditions.

\begin{figure}[htbp]
    \centering
    \begin{minipage}[b]{0.45\textwidth}
        \centering
        \includegraphics[width=\textwidth]{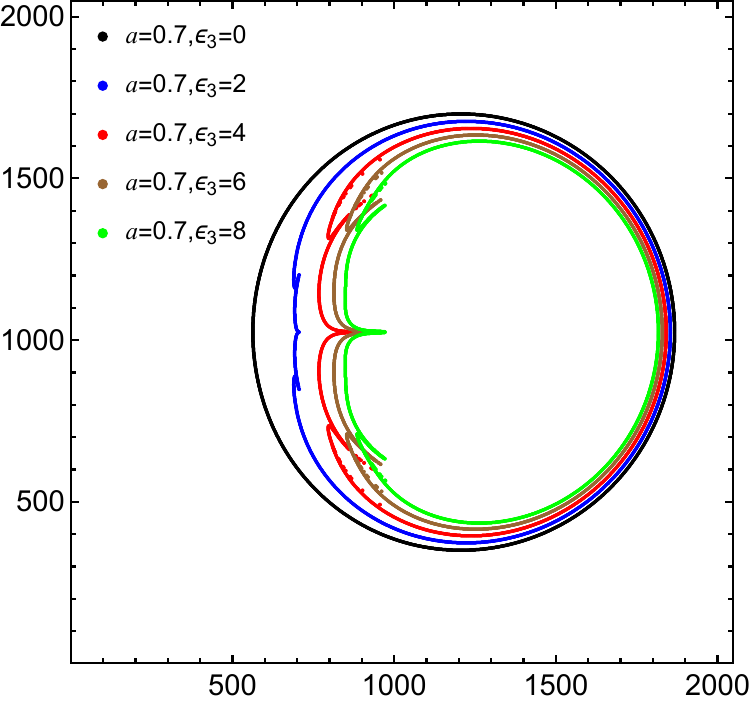}
        \caption*{(a)} 
    \end{minipage}
    \begin{minipage}[b]{0.45\textwidth}
        \centering
        \includegraphics[width=\textwidth]{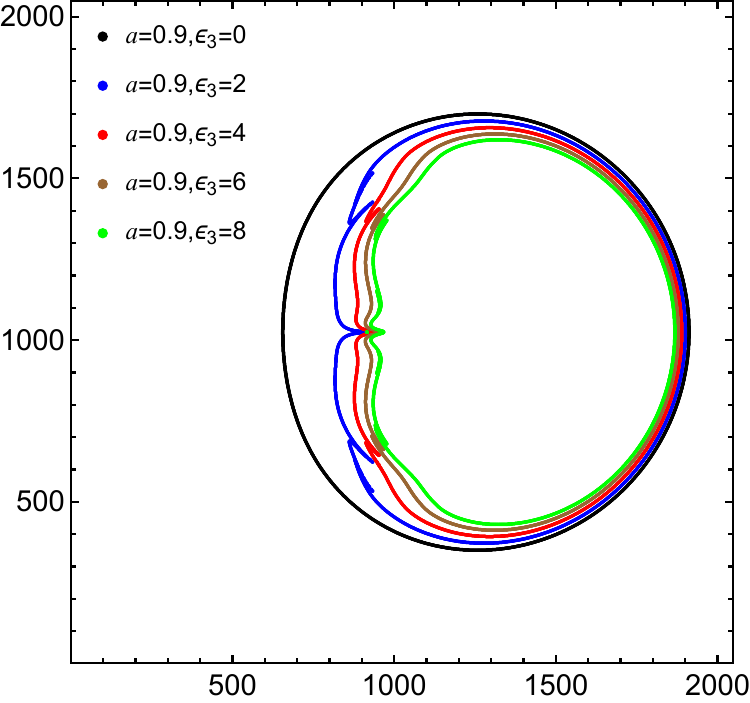}
        \caption*{(b)}
    \end{minipage}
    \caption{The evolution of the critical curve of the JP black hole shadow with increasing the parameter $\epsilon_3$ is illustrated, focusing on positive values where the event horizon remains non-closed. Panel (a) represents the JP black hole with $a = 0.7$, while Panel (b) corresponds to the case with $a = 0.9$.}
    \label{variationccwithepsilon}
\end{figure}

Building on this foundation, a comprehensive numerical analysis investigates the evolution of the critical curve of the JP black hole shadow as the parameter $\epsilon_3$ increases under conditions where the event horizon remains non-closed. The corresponding results are illustrated in Fig. \ref{variationccwithepsilon}.
It is demonstrated that for the JP black hole with spin parameters $a = 0.7$ and $a = 0.9$, the evolution of the critical curve of the shadow exhibits a consistent trend as $\epsilon_3$ increases. Specifically, the critical curve becomes increasingly concave relative to the corresponding Kerr black hole shadow. Notably, near the concave region, the critical curve transitions from a smooth profile to exhibit distinctive and unconventional irregularities, revealing novel structural properties of the JP black hole shadow.

To comprehensively investigate the distinctive non-smooth features of the critical curve of the JP black hole shadow with a non-closed event horizon as the parameter $\epsilon_3$ increases, it is essential to analyze the properties of photon trajectories that traverse the JP black hole spacetime and converge on the concave and non-smooth region of the critical curve. For this purpose, several closely spaced points within the non-smooth region of the critical curve are selected. The backward ray-tracing numerical method is utilized to compute the trajectories of photons associated with these points as they propagate through the JP black hole spacetime. Based on these calculations, a systematic analysis of the photon trajectories converging on the non-smooth region of the critical curve is conducted to elucidate the underlying physical mechanisms of the non-smooth features of the critical curve. 
\begin{figure}[htbp]
    \centering
    \begin{minipage}[b]{0.55\textwidth}
        \centering
        \includegraphics[width=\textwidth]{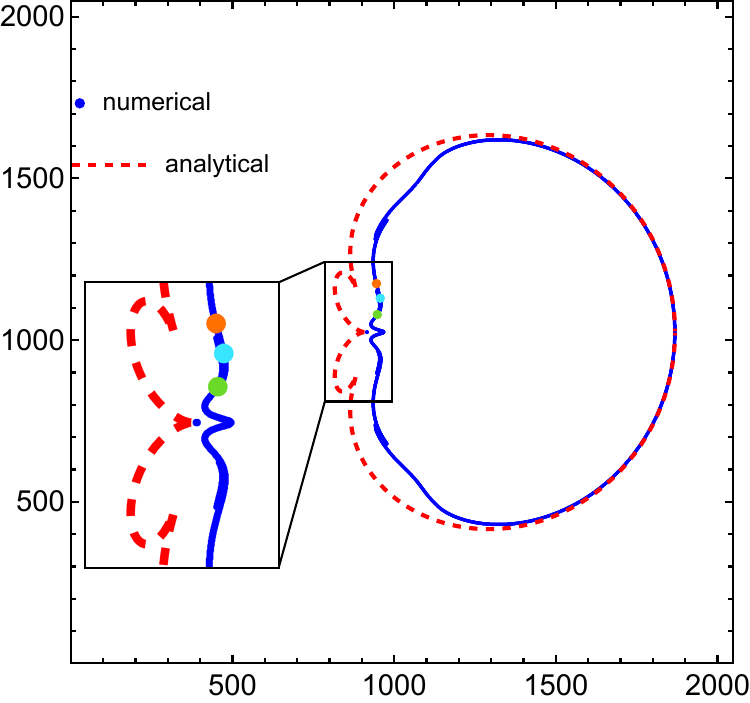}
        \caption*{(a)} 
    \end{minipage}

    \vspace{1em}

    \begin{minipage}[b]{1\textwidth}
        \centering
        \includegraphics[width=\textwidth]{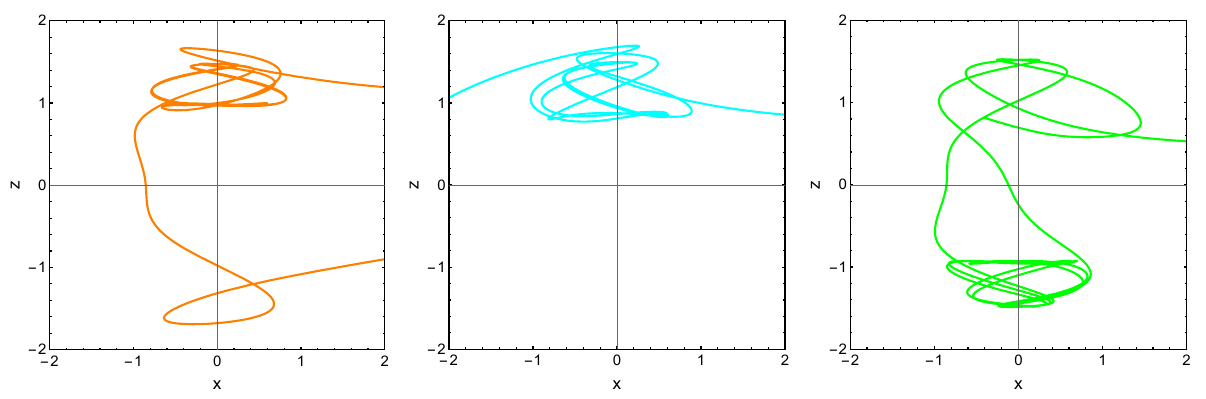}
        \caption*{(b)}
    \end{minipage}
    \caption{(a) For the JP black hole shadow with $a = 0.9$ and $\epsilon_3 = 8$, three closely spaced reference points in the non-smooth region of the critical curve are marked in orange, cyan, and green, respectively. (b) The projection of photon trajectories onto the $xz$-plane, originating from the three reference points and propagating through the JP black hole spacetime, is illustrated. The colors of the trajectories correspond to those of the respective reference points.}
    \label{chasoofphotontraj}
\end{figure}
Figure \ref{chasoofphotontraj} (a) illustrates three closely spaced reference points, marked in orange, cyan, and green, situated within the non-smooth region of the critical curve for the shadow of a JP black hole with the spin parameter $a = 0.9$ and $\epsilon_3 = 8$. The photon trajectories projected onto the $xz$-plane, originating from the three reference points in the non-smooth region of the critical curve and traversing the JP black hole spacetime, are shown in Figure \ref{chasoofphotontraj} (b). The colors of the trajectories correspond to those of the respective reference points. These trajectories exhibit pronounced complexity, particularly within the concave and non-smooth region of the critical curve. Specifically, for a fixed incident direction, the paths of photons through the JP black hole spacetime and their emergent directions are highly unpredictable. This unpredictability indicates that when employing the backward ray-tracing method to numerically compute the JP black hole shadow, even infinitesimal changes in the initial conditions for photon trajectories at the critical curve can result in highly unpredictable paths, as governed by the Hamilton canonical equations. This characteristic unpredictability aligns with the defining attributes of chaotic phenomena. Therefore, it can be inferred that photon trajectories associated with the concave and non-smooth region of the critical curve exhibit chaotic behavior as they propagate through the JP black hole spacetime. Moreover, as shown in Figure \ref{relationanalynumernonclosed}, the discrepancies between the critical curves derived using the approximate analytical method and those obtained through the numerical computation are predominantly localized in this region. This observation underscores that the chaotic behavior of photon trajectories is the primary factor responsible for the failure of the approximate analytical method to accurately predict the JP black hole shadow when the event horizon is not closed.

\section{Conclusions}\label{Summary}

The no-hair theorem asserts that stationary black holes are fully described by three parameters: mass, charge, and angular momentum. Astrophysical black holes, characterized by the mass and angular momentum, are effectively modeled by Kerr black holes, which rigorously conform to this theorem. However, the presence of surrounding celestial bodies and matter fields creates non-vacuum environments, leading to deviations in the spacetime metrics of astrophysical black holes from the Kerr solution. Consequently, the applicability of the no-hair theorem to astrophysical black holes requires further investigation. Photon circular orbits and ISCOs are critical indicators for testing the no-hair theorem via electromagnetic spectra, as their configurations are highly sensitive to deviations from the Kerr metric. Additionally, black hole shadows, which are strongly correlated with photon circular orbits, provide a complementary observational tool for evaluating the theorem. To rigorously test the no-hair theorem for astrophysical black holes, it is essential to employ appropriate spacetime metrics that systematically incorporate deviations from the Kerr metric.

The JP metric represents a viable model for testing the no-hair theorem, as the spacetime region outside the event horizon of a JP black hole remains regular, enabling detailed investigation of black hole shadow properties. The JP metric deviates from the Kerr metric through the inclusion of the deviation parameter $\epsilon_3$, which alters its Petrov classification from Petrov-type D to Petrov-type I. Therefore, the Carter constant is undefined, and the Hamilton-Jacobi equation for photon motion loses strict separability. However, under the limiting condition $\epsilon_3 \to 0$, approximate separability is preserved, allowing the JP black hole shadow to be determined using approximate analytical techniques. Additionally, the numerical method employing backward ray-tracing provide precise calculations of JP black hole shadows. Comparing the results from the approximate analytical method with those obtained through the numerical computation enables an evaluation of the validity and applicability of the approximate analytical approach in calculating JP black hole shadows.

For JP black holes with a closed event horizon, the critical curve of the JP black hole shadow was initially derived using an approximate analytical method. The analysis revealed that the critical curve of JP black hole shadow aligns precisely with that of the Kerr black hole shadow when the function $\tilde{\mathcal{I}} (r, \theta)$ is excluded from the radial potential $\mathcal{R} (r)$ of the photon. However, when $\tilde{\mathcal{I}} (r, \theta)$ is included in $\mathcal{R} (r)$, significant derivations in the critical curve of the JP black hole shadow from that of the Kerr black hole shadow arise. Specifically, the critical curve of the JP black hole shadow expands relative to the Kerr black hole shadow and adopts an elliptical configuration with its major axis oriented along the $x$-axis. We also employed the numerical method to determine the critical curve of the JP black hole shadow. Comparing the critical curve obtained through the numerical calculation with that derived using the approximate analytical method, we have found that when the function $\tilde{\mathcal{I}}(r, \theta)$ is included in the radial potential $\mathcal{R} (r)$, the critical curve of the JP black hole shadow derived using the approximate analytical method aligns perfectly with that obtained through the numerical calculation. These results revealed that the inclusion of $\tilde{\mathcal{I}}(r, \theta)$ in the radial potential $\mathcal{R}(r)$ is essential for accurately deriving the critical curve of the JP black hole shadow using the approximate analytical method. Additionally, for JP black holes with a closed event horizon, the approximate analytical method has been shown to be a reliable alternative to the numerical computation for precisely determining the critical curve of the JP black hole shadow. Subsequently, we analyzed the relationship between the critical curve of the JP black hole shadow and the deviation parameter $\epsilon_3$, elucidating the variations in the behavior of the critical curve as a function of $\epsilon_3$. We have observed that for $\epsilon_3 < 0$, the critical curve of the JP black hole shadow expands outward relative to the Kerr black hole shadow and progressively adopts an elliptical shape with its major axis oriented along the $x$-axis as the absolute value of $\epsilon_3$ increases. Conversely, for $\epsilon_3 > 0$, the critical curve contracts inward compared to that of the Kerr black hole and exhibits a concave distortion near the region around $y = 200$ as $\epsilon_3$ increases. However, these variations remain relatively minor for $\epsilon_3 > 0$, and the critical curve closely resembles the Kerr black hole shadow.

For JP black holes with non-closed event horizons, we have obtained that the critical curve of the JP black hole shadow derived using approximate analytical approach deviates significantly from that obtained through the numerical calculation. Therefore, the approximate analytical approach cannot reliably substitute the numerical method for calculating JP black hole shadows. Analysis of the variation in the critical curve of the JP black hole shadow with respect to $\epsilon_3$ revealed that as $\epsilon_3$ increases, the critical curve becomes increasingly concave relative to the Kerr black hole shadow, with its contour exhibiting a noticeable loss of smoothness. Investigation of photon trajectories in the non-smooth region of the critical curve uncovered the emergence of chaotic dynamics as photons propagate through the JP black hole spacetime. These chaotic photon trajectories have been identified as the primary reason for the failure of the approximate analytical method in accurately computing the shadows of JP black holes with non-closed event horizons.

\section*{Acknowledgments}
We would like to thank Minyong Guo for useful discussions. This work is supported by the National Natural Science Foundation of China with Grant No. 12105015, the Guangdong Basic and Applied Basic Research Foundation with Grant No. 2023A1515012737, and the Talents Introduction Foundation of Beijing Normal University with Grant No. 111032109.

\bibliographystyle{utphys}

\bibliography{JP}

\providecommand{\href}[2]{#2}\begingroup\raggedright\begin{thebibliography}{10}

\bibitem{EventHorizonTelescope:2019dse}
{\bfseries Event Horizon Telescope} Collaboration, K.~Akiyama {\em et~al.},
  ``{First M87 Event Horizon Telescope Results. I. The Shadow of the
  Supermassive Black Hole},''
  \href{http://dx.doi.org/10.3847/2041-8213/ab0ec7}{{\em Astrophys. J. Lett.}
  {\bfseries 875} (2019) L1}, \href{http://arxiv.org/abs/1906.11238}{{\ttfamily
  arXiv:1906.11238 [astro-ph.GA]}}.

\bibitem{EventHorizonTelescope:2019uob}
{\bfseries Event Horizon Telescope} Collaboration, K.~Akiyama {\em et~al.},
  ``{First M87 Event Horizon Telescope Results. II. Array and
  Instrumentation},'' \href{http://dx.doi.org/10.3847/2041-8213/ab0c96}{{\em
  Astrophys. J. Lett.} {\bfseries 875} no.~1, (2019) L2},
  \href{http://arxiv.org/abs/1906.11239}{{\ttfamily arXiv:1906.11239
  [astro-ph.IM]}}.

\bibitem{EventHorizonTelescope:2019jan}
{\bfseries Event Horizon Telescope} Collaboration, K.~Akiyama {\em et~al.},
  ``{First M87 Event Horizon Telescope Results. III. Data Processing and
  Calibration},'' \href{http://dx.doi.org/10.3847/2041-8213/ab0c57}{{\em
  Astrophys. J. Lett.} {\bfseries 875} no.~1, (2019) L3},
  \href{http://arxiv.org/abs/1906.11240}{{\ttfamily arXiv:1906.11240
  [astro-ph.GA]}}.

\bibitem{EventHorizonTelescope:2019ths}
{\bfseries Event Horizon Telescope} Collaboration, K.~Akiyama {\em et~al.},
  ``{First M87 Event Horizon Telescope Results. IV. Imaging the Central
  Supermassive Black Hole},''
  \href{http://dx.doi.org/10.3847/2041-8213/ab0e85}{{\em Astrophys. J. Lett.}
  {\bfseries 875} no.~1, (2019) L4},
  \href{http://arxiv.org/abs/1906.11241}{{\ttfamily arXiv:1906.11241
  [astro-ph.GA]}}.

\bibitem{EventHorizonTelescope:2019pgp}
{\bfseries Event Horizon Telescope} Collaboration, K.~Akiyama {\em et~al.},
  ``{First M87 Event Horizon Telescope Results. V. Physical Origin of the
  Asymmetric Ring},'' \href{http://dx.doi.org/10.3847/2041-8213/ab0f43}{{\em
  Astrophys. J. Lett.} {\bfseries 875} no.~1, (2019) L5},
  \href{http://arxiv.org/abs/1906.11242}{{\ttfamily arXiv:1906.11242
  [astro-ph.GA]}}.

\bibitem{EventHorizonTelescope:2019ggy}
{\bfseries Event Horizon Telescope} Collaboration, K.~Akiyama {\em et~al.},
  ``{First M87 Event Horizon Telescope Results. VI. The Shadow and Mass of the
  Central Black Hole},'' \href{http://dx.doi.org/10.3847/2041-8213/ab1141}{{\em
  Astrophys. J. Lett.} {\bfseries 875} no.~1, (2019) L6},
  \href{http://arxiv.org/abs/1906.11243}{{\ttfamily arXiv:1906.11243
  [astro-ph.GA]}}.

\bibitem{EventHorizonTelescope:2021bee}
{\bfseries Event Horizon Telescope} Collaboration, K.~Akiyama {\em et~al.},
  ``{First M87 Event Horizon Telescope Results. VII. Polarization of the
  Ring},'' \href{http://dx.doi.org/10.3847/2041-8213/abe71d}{{\em Astrophys. J.
  Lett.} {\bfseries 910} no.~1, (2021) L12},
  \href{http://arxiv.org/abs/2105.01169}{{\ttfamily arXiv:2105.01169
  [astro-ph.HE]}}.

\bibitem{1983mtbh.book.....C}
S.~{Chandrasekhar}, {\em {The mathematical theory of black holes}}.
\newblock 1983.

\bibitem{Cunha:2018acu}
P.~V.~P. Cunha and C.~A.~R. Herdeiro, ``{Shadows and strong gravitational
  lensing: a brief review},''
  \href{http://dx.doi.org/10.1007/s10714-018-2361-9}{{\em Gen. Rel. Grav.}
  {\bfseries 50} no.~4, (2018) 42},
  \href{http://arxiv.org/abs/1801.00860}{{\ttfamily arXiv:1801.00860 [gr-qc]}}.

\bibitem{Perlick:2021aok}
V.~Perlick and O.~Y. Tsupko, ``{Calculating black hole shadows: Review of
  analytical studies},''
  \href{http://dx.doi.org/10.1016/j.physrep.2021.10.004}{{\em Phys. Rept.}
  {\bfseries 947} (2022) 1--39},
  \href{http://arxiv.org/abs/2105.07101}{{\ttfamily arXiv:2105.07101 [gr-qc]}}.

\bibitem{Wang:2023nwd}
X.~Wang, Y.~Hou, and M.~Guo, ``{How different are shadows of compact objects
  with and without horizons?},''
  \href{http://dx.doi.org/10.1088/1475-7516/2023/05/036}{{\em JCAP} {\bfseries
  05} (2023) 036}, \href{http://arxiv.org/abs/2301.04851}{{\ttfamily
  arXiv:2301.04851 [gr-qc]}}.

\bibitem{Wang:2024uda}
X.~Wang, X.~Wang, H.-Q. Zhang, and M.~Guo, ``{Is a photon ring invariably a
  closed structure?},''
  \href{http://dx.doi.org/10.1140/epjc/s10052-024-13527-6}{{\em Eur. Phys. J.
  C} {\bfseries 84} no.~11, (2024) 1168},
  \href{http://arxiv.org/abs/2405.05011}{{\ttfamily arXiv:2405.05011 [gr-qc]}}.

\bibitem{Wei:2013kza}
S.-W. Wei and Y.-X. Liu, ``{Observing the shadow of
  Einstein-Maxwell-Dilaton-Axion black hole},''
  \href{http://dx.doi.org/10.1088/1475-7516/2013/11/063}{{\em JCAP} {\bfseries
  11} (2013) 063}, \href{http://arxiv.org/abs/1311.4251}{{\ttfamily
  arXiv:1311.4251 [gr-qc]}}.

\bibitem{Grenzebach:2014fha}
A.~Grenzebach, V.~Perlick, and C.~L\"ammerzahl, ``{Photon Regions and Shadows
  of Kerr-Newman-NUT Black Holes with a Cosmological Constant},''
  \href{http://dx.doi.org/10.1103/PhysRevD.89.124004}{{\em Phys. Rev. D}
  {\bfseries 89} no.~12, (2014) 124004},
  \href{http://arxiv.org/abs/1403.5234}{{\ttfamily arXiv:1403.5234 [gr-qc]}}.

\bibitem{Cunha:2016bpi}
P.~V.~P. Cunha, C.~A.~R. Herdeiro, E.~Radu, and H.~F. Runarsson, ``{Shadows of
  Kerr black holes with and without scalar hair},''
  \href{http://dx.doi.org/10.1142/S0218271816410212}{{\em Int. J. Mod. Phys. D}
  {\bfseries 25} no.~09, (2016) 1641021},
  \href{http://arxiv.org/abs/1605.08293}{{\ttfamily arXiv:1605.08293 [gr-qc]}}.

\bibitem{Cunha:2016wzk}
P.~V.~P. Cunha, C.~A.~R. Herdeiro, B.~Kleihaus, J.~Kunz, and E.~Radu,
  ``{Shadows of
  Einstein\textendash{}dilaton\textendash{}Gauss\textendash{}Bonnet black
  holes},'' \href{http://dx.doi.org/10.1016/j.physletb.2017.03.020}{{\em Phys.
  Lett. B} {\bfseries 768} (2017) 373--379},
  \href{http://arxiv.org/abs/1701.00079}{{\ttfamily arXiv:1701.00079 [gr-qc]}}.

\bibitem{Guo:2020zmf}
M.~Guo and P.-C. Li, ``{Innermost stable circular orbit and shadow of the $4D$
  Einstein\textendash{}Gauss\textendash{}Bonnet black hole},''
  \href{http://dx.doi.org/10.1140/epjc/s10052-020-8164-7}{{\em Eur. Phys. J. C}
  {\bfseries 80} no.~6, (2020) 588},
  \href{http://arxiv.org/abs/2003.02523}{{\ttfamily arXiv:2003.02523 [gr-qc]}}.

\bibitem{Chen:2023trn}
Y.~Chen, P.~Wang, H.~Wu, and H.~Yang, ``{Gravitational lensing by Born-Infeld
  naked singularities},''
  \href{http://dx.doi.org/10.1103/PhysRevD.109.084014}{{\em Phys. Rev. D}
  {\bfseries 109} no.~8, (2024) 084014},
  \href{http://arxiv.org/abs/2305.17411}{{\ttfamily arXiv:2305.17411 [gr-qc]}}.

\bibitem{Hou:2021okc}
Y.~Hou, M.~Guo, and B.~Chen, ``{Revisiting the shadow of braneworld black
  holes},'' \href{http://dx.doi.org/10.1103/PhysRevD.104.024001}{{\em Phys.
  Rev. D} {\bfseries 104} no.~2, (2021) 024001},
  \href{http://arxiv.org/abs/2103.04369}{{\ttfamily arXiv:2103.04369 [gr-qc]}}.

\bibitem{Wei:2020ght}
S.-W. Wei and Y.-X. Liu, ``{Testing the nature of Gauss-Bonnet gravity by
  four-dimensional rotating black hole shadow},''
  \href{http://dx.doi.org/10.1140/epjp/s13360-021-01398-9}{{\em Eur. Phys. J.
  Plus} {\bfseries 136} no.~4, (2021) 436},
  \href{http://arxiv.org/abs/2003.07769}{{\ttfamily arXiv:2003.07769 [gr-qc]}}.

\bibitem{Ovgun:2020gjz}
A.~\"Ovg\"un and I.~Sakall\i{}, ``{Testing generalized
  Einstein\textendash{}Cartan\textendash{}Kibble\textendash{}Sciama gravity
  using weak deflection angle and shadow cast},''
  \href{http://dx.doi.org/10.1088/1361-6382/abb579}{{\em Class. Quant. Grav.}
  {\bfseries 37} no.~22, (2020) 225003},
  \href{http://arxiv.org/abs/2005.00982}{{\ttfamily arXiv:2005.00982 [gr-qc]}}.

\bibitem{Kuang:2022ojj}
X.-M. Kuang, Z.-Y. Tang, B.~Wang, and A.~Wang, ``{Constraining a modified
  gravity theory in strong gravitational lensing and black hole shadow
  observations},'' \href{http://dx.doi.org/10.1103/PhysRevD.106.064012}{{\em
  Phys. Rev. D} {\bfseries 106} no.~6, (2022) 064012},
  \href{http://arxiv.org/abs/2206.05878}{{\ttfamily arXiv:2206.05878 [gr-qc]}}.

\bibitem{Haroon:2018ryd}
S.~Haroon, M.~Jamil, K.~Jusufi, K.~Lin, and R.~B. Mann, ``{Shadow and
  Deflection Angle of Rotating Black Holes in Perfect Fluid Dark Matter with a
  Cosmological Constant},''
  \href{http://dx.doi.org/10.1103/PhysRevD.99.044015}{{\em Phys. Rev. D}
  {\bfseries 99} no.~4, (2019) 044015},
  \href{http://arxiv.org/abs/1810.04103}{{\ttfamily arXiv:1810.04103 [gr-qc]}}.

\bibitem{Konoplya:2019sns}
R.~A. Konoplya, ``{Shadow of a black hole surrounded by dark matter},''
  \href{http://dx.doi.org/10.1016/j.physletb.2019.05.043}{{\em Phys. Lett. B}
  {\bfseries 795} (2019) 1--6},
  \href{http://arxiv.org/abs/1905.00064}{{\ttfamily arXiv:1905.00064 [gr-qc]}}.

\bibitem{Chen:2019fsq}
Y.~Chen, J.~Shu, X.~Xue, Q.~Yuan, and Y.~Zhao, ``{Probing Axions with Event
  Horizon Telescope Polarimetric Measurements},''
  \href{http://dx.doi.org/10.1103/PhysRevLett.124.061102}{{\em Phys. Rev.
  Lett.} {\bfseries 124} no.~6, (2020) 061102},
  \href{http://arxiv.org/abs/1905.02213}{{\ttfamily arXiv:1905.02213
  [hep-ph]}}.

\bibitem{Chen:2021lvo}
Y.~Chen, Y.~Liu, R.-S. Lu, Y.~Mizuno, J.~Shu, X.~Xue, Q.~Yuan, and Y.~Zhao,
  ``{Stringent axion constraints with Event Horizon Telescope polarimetric
  measurements of M87*},''
  \href{http://dx.doi.org/10.1038/s41550-022-01620-3}{{\em Nature Astron.}
  {\bfseries 6} no.~5, (2022) 592--598},
  \href{http://arxiv.org/abs/2105.04572}{{\ttfamily arXiv:2105.04572
  [hep-ph]}}.

\bibitem{Faraji:2024ein}
S.~Faraji and J.~a.~L. Rosa, ``{Effect of dark matter on the shadow of a
  distorted and deformed compact object},''
  \href{http://arxiv.org/abs/2403.02597}{{\ttfamily arXiv:2403.02597
  [astro-ph.HE]}}.

\bibitem{Zhang:2024hjr}
C.~Zhang, G.~Fu, and C.~Zhang, ``{Rotating galactic black holes},''
  \href{http://arxiv.org/abs/2403.19933}{{\ttfamily arXiv:2403.19933 [gr-qc]}}.

\bibitem{Zeng:2024ptv}
X.-X. Zeng, L.-F. Li, P.~Li, B.~Liang, and P.~Xu, ``{Holographic images of a
  charged black hole in Lorentz symmetry breaking massive gravity},''
  \href{http://dx.doi.org/10.1007/s11433-024-2526-4}{{\em Sci. China Phys.
  Mech. Astron.} {\bfseries 68} no.~2, (2025) 220412},
  \href{http://arxiv.org/abs/2411.12528}{{\ttfamily arXiv:2411.12528 [gr-qc]}}.

\bibitem{newman1965metric}
E.~T. Newman, E.~Couch, K.~Chinnapared, A.~Exton, A.~Prakash, and R.~Torrence,
  ``Metric of a rotating, charged mass,'' {\em Journal of mathematical physics}
  {\bfseries 6} no.~6, (1965) 918--919.

\bibitem{Ryan:1997hg}
F.~D. Ryan, ``{Accuracy of estimating the multipole moments of a massive body
  from the gravitational waves of a binary inspiral},''
  \href{http://dx.doi.org/10.1103/PhysRevD.56.1845}{{\em Phys. Rev. D}
  {\bfseries 56} (1997) 1845--1855}.

\bibitem{Gair:2011ym}
J.~Gair and N.~Yunes, ``{Approximate Waveforms for Extreme-Mass-Ratio Inspirals
  in Modified Gravity Spacetimes},''
  \href{http://dx.doi.org/10.1103/PhysRevD.84.064016}{{\em Phys. Rev. D}
  {\bfseries 84} (2011) 064016},
  \href{http://arxiv.org/abs/1106.6313}{{\ttfamily arXiv:1106.6313 [gr-qc]}}.

\bibitem{Johannsen:2010xs}
T.~Johannsen and D.~Psaltis, ``{Testing the No-Hair Theorem with Observations
  in the Electromagnetic Spectrum: I. Properties of a Quasi-Kerr Spacetime},''
  \href{http://dx.doi.org/10.1088/0004-637X/716/1/187}{{\em Astrophys. J.}
  {\bfseries 716} (2010) 187--197},
  \href{http://arxiv.org/abs/1003.3415}{{\ttfamily arXiv:1003.3415
  [astro-ph.HE]}}.

\bibitem{Johannsen:2010ru}
T.~Johannsen and D.~Psaltis, ``{Testing the No-Hair Theorem with Observations
  in the Electromagnetic Spectrum: II. Black-Hole Images},''
  \href{http://dx.doi.org/10.1088/0004-637X/718/1/446}{{\em Astrophys. J.}
  {\bfseries 718} (2010) 446--454},
  \href{http://arxiv.org/abs/1005.1931}{{\ttfamily arXiv:1005.1931
  [astro-ph.HE]}}.

\bibitem{Johannsen:2010bi}
T.~Johannsen and D.~Psaltis, ``{Testing the No-Hair Theorem with Observations
  in the Electromagnetic Spectrum. III. Quasi-Periodic Variability},''
  \href{http://dx.doi.org/10.1088/0004-637X/726/1/11}{{\em Astrophys. J.}
  {\bfseries 726} (2011) 11}, \href{http://arxiv.org/abs/1010.1000}{{\ttfamily
  arXiv:1010.1000 [astro-ph.HE]}}.

\bibitem{Collins:2004ex}
N.~A. Collins and S.~A. Hughes, ``{Towards a formalism for mapping the
  space-times of massive compact objects: Bumpy black holes and their
  orbits},'' \href{http://dx.doi.org/10.1103/PhysRevD.69.124022}{{\em Phys.
  Rev. D} {\bfseries 69} (2004) 124022},
  \href{http://arxiv.org/abs/gr-qc/0402063}{{\ttfamily arXiv:gr-qc/0402063}}.

\bibitem{Vigeland:2009pr}
S.~J. Vigeland and S.~A. Hughes, ``{Spacetime and orbits of bumpy black
  holes},'' \href{http://dx.doi.org/10.1103/PhysRevD.81.024030}{{\em Phys. Rev.
  D} {\bfseries 81} (2010) 024030},
  \href{http://arxiv.org/abs/0911.1756}{{\ttfamily arXiv:0911.1756 [gr-qc]}}.

\bibitem{Glampedakis:2005cf}
K.~Glampedakis and S.~Babak, ``{Mapping spacetimes with LISA: Inspiral of a
  test-body in a `quasi-Kerr' field},''
  \href{http://dx.doi.org/10.1088/0264-9381/23/12/013}{{\em Class. Quant.
  Grav.} {\bfseries 23} (2006) 4167--4188},
  \href{http://arxiv.org/abs/gr-qc/0510057}{{\ttfamily arXiv:gr-qc/0510057}}.

\bibitem{Mankot1992GeneralizationsOT}
V.~S. Mankot, ``Generalizations of the kerr and kerr-newman metrics possessing
  an arbitrary set of mass-multipole moments,'' {\em Classical and Quantum
  Gravity} {\bfseries 9} (1992) 2477 -- 2487.
  \url{https://api.semanticscholar.org/CorpusID:121051772}.

\bibitem{Vigeland:2011ji}
S.~Vigeland, N.~Yunes, and L.~Stein, ``{Bumpy Black Holes in Alternate Theories
  of Gravity},'' \href{http://dx.doi.org/10.1103/PhysRevD.83.104027}{{\em Phys.
  Rev. D} {\bfseries 83} (2011) 104027},
  \href{http://arxiv.org/abs/1102.3706}{{\ttfamily arXiv:1102.3706 [gr-qc]}}.

\bibitem{Johannsen:2013szh}
T.~Johannsen, ``{Regular Black Hole Metric with Three Constants of Motion},''
  \href{http://dx.doi.org/10.1103/PhysRevD.88.044002}{{\em Phys. Rev. D}
  {\bfseries 88} no.~4, (2013) 044002},
  \href{http://arxiv.org/abs/1501.02809}{{\ttfamily arXiv:1501.02809 [gr-qc]}}.

\bibitem{Johannsen:2013rqa}
T.~Johannsen, ``{Systematic Study of Event Horizons and Pathologies of
  Parametrically Deformed Kerr Spacetimes},''
  \href{http://dx.doi.org/10.1103/PhysRevD.87.124017}{{\em Phys. Rev. D}
  {\bfseries 87} no.~12, (2013) 124017},
  \href{http://arxiv.org/abs/1304.7786}{{\ttfamily arXiv:1304.7786 [gr-qc]}}.

\bibitem{Johannsen:2011dh}
T.~Johannsen and D.~Psaltis, ``{A Metric for Rapidly Spinning Black Holes
  Suitable for Strong-Field Tests of the No-Hair Theorem},''
  \href{http://dx.doi.org/10.1103/PhysRevD.83.124015}{{\em Phys. Rev. D}
  {\bfseries 83} (2011) 124015},
  \href{http://arxiv.org/abs/1105.3191}{{\ttfamily arXiv:1105.3191 [gr-qc]}}.

\bibitem{Atamurotov:2013sca}
F.~Atamurotov, A.~Abdujabbarov, and B.~Ahmedov, ``{Shadow of rotating non-Kerr
  black hole},'' \href{http://dx.doi.org/10.1103/PhysRevD.88.064004}{{\em Phys.
  Rev. D} {\bfseries 88} no.~6, (2013) 064004}.

\bibitem{Bambhaniya:2021ybs}
P.~Bambhaniya, D.~Dey, A.~B. Joshi, P.~S. Joshi, D.~N. Solanki, and A.~Mehta,
  ``{Shadows and negative precession in non-Kerr spacetime},''
  \href{http://dx.doi.org/10.1103/PhysRevD.103.084005}{{\em Phys. Rev. D}
  {\bfseries 103} no.~8, (2021) 084005},
  \href{http://arxiv.org/abs/2101.03865}{{\ttfamily arXiv:2101.03865 [gr-qc]}}.

\bibitem{Khodadi:2021gbc}
M.~Khodadi, G.~Lambiase, and D.~F. Mota, ``{No-hair theorem in the wake of
  Event Horizon Telescope},''
  \href{http://dx.doi.org/10.1088/1475-7516/2021/09/028}{{\em JCAP} {\bfseries
  09} (2021) 028}, \href{http://arxiv.org/abs/2107.00834}{{\ttfamily
  arXiv:2107.00834 [gr-qc]}}.

\bibitem{Li:2014fza}
Z.~Li, L.~Kong, and C.~Bambi, ``{Testing the nature of the supermassive black
  hole candidate in SgrA* with light curves and images of hot spots},''
  \href{http://dx.doi.org/10.1088/0004-637X/787/2/152}{{\em Astrophys. J.}
  {\bfseries 787} (2014) 152}, \href{http://arxiv.org/abs/1401.1282}{{\ttfamily
  arXiv:1401.1282 [gr-qc]}}.

\bibitem{Liu:2014awa}
D.~Liu, Z.~Li, and C.~Bambi, ``{Testing a class of non-Kerr metrics with hot
  spots orbiting SgrA*},''
  \href{http://dx.doi.org/10.1088/1475-7516/2015/01/020}{{\em JCAP} {\bfseries
  01} (2015) 020}, \href{http://arxiv.org/abs/1411.2329}{{\ttfamily
  arXiv:1411.2329 [gr-qc]}}.

\bibitem{Bambi:2012tg}
C.~Bambi, ``{A code to compute the emission of thin accretion disks in non-Kerr
  space-times and test the nature of black hole candidates},''
  \href{http://dx.doi.org/10.1088/0004-637X/761/2/174}{{\em Astrophys. J.}
  {\bfseries 761} (2012) 174}, \href{http://arxiv.org/abs/1210.5679}{{\ttfamily
  arXiv:1210.5679 [gr-qc]}}.

\bibitem{Kong:2014wha}
L.~Kong, Z.~Li, and C.~Bambi, ``{Constraints on the spacetime geometry around
  10 stellar-mass black hole candidates from the disk's thermal spectrum},''
  \href{http://dx.doi.org/10.1088/0004-637X/797/2/78}{{\em Astrophys. J.}
  {\bfseries 797} no.~2, (2014) 78},
  \href{http://arxiv.org/abs/1405.1508}{{\ttfamily arXiv:1405.1508 [gr-qc]}}.

\bibitem{Li:2012ra}
Z.~Li and C.~Bambi, ``{Super-spinning compact objects generated by thick
  accretion disks},''
  \href{http://dx.doi.org/10.1088/1475-7516/2013/03/031}{{\em JCAP} {\bfseries
  03} (2013) 031}, \href{http://arxiv.org/abs/1212.5848}{{\ttfamily
  arXiv:1212.5848 [gr-qc]}}.

\bibitem{John:2019rhj}
A.~John and C.~Stevens, ``{Bondi accretion in the spherically symmetric
  Johannsen\textendash{}Psaltis spacetime},''
  \href{http://dx.doi.org/10.1140/epjc/s10052-019-7481-1}{{\em Eur. Phys. J. C}
  {\bfseries 79} no.~11, (2019) 962},
  \href{http://arxiv.org/abs/1903.01958}{{\ttfamily arXiv:1903.01958 [gr-qc]}}.

\bibitem{Rahim:2018ruj}
R.~Rahim and K.~Saifullah, ``{Charging the Johannsen\textendash{}Psaltis
  spacetime},'' \href{http://dx.doi.org/10.1016/j.aop.2019.03.007}{{\em Annals
  Phys.} {\bfseries 405} (2019) 220--233},
  \href{http://arxiv.org/abs/1810.10504}{{\ttfamily arXiv:1810.10504 [gr-qc]}}.

\bibitem{Carter:1968rr}
B.~Carter, ``{Global structure of the Kerr family of gravitational fields},''
  \href{http://dx.doi.org/10.1103/PhysRev.174.1559}{{\em Phys. Rev.} {\bfseries
  174} (1968) 1559--1571}.

\bibitem{Johannsen:2013vgc}
T.~Johannsen, ``{Photon Rings around Kerr and Kerr-like Black Holes},''
  \href{http://dx.doi.org/10.1088/0004-637X/777/2/170}{{\em Astrophys. J.}
  {\bfseries 777} (2013) 170},
  \href{http://arxiv.org/abs/1501.02814}{{\ttfamily arXiv:1501.02814
  [astro-ph.HE]}}.

\bibitem{Glampedakis:2018blj}
K.~Glampedakis and G.~Pappas, ``{Modification of photon trapping orbits as a
  diagnostic of non-Kerr spacetimes},''
  \href{http://dx.doi.org/10.1103/PhysRevD.99.124041}{{\em Phys. Rev. D}
  {\bfseries 99} no.~12, (2019) 124041},
  \href{http://arxiv.org/abs/1806.09333}{{\ttfamily arXiv:1806.09333 [gr-qc]}}.

\bibitem{Younsi:2021dxe}
Z.~Younsi, D.~Psaltis, and F.~\"Ozel, ``{Black Hole Images as Tests of General
  Relativity: Effects of Spacetime Geometry},''
  \href{http://dx.doi.org/10.3847/1538-4357/aca58a}{{\em Astrophys. J.}
  {\bfseries 942} no.~1, (2023) 47},
  \href{http://arxiv.org/abs/2111.01752}{{\ttfamily arXiv:2111.01752
  [astro-ph.HE]}}.

\bibitem{Hu:2020usx}
Z.~Hu, Z.~Zhong, P.-C. Li, M.~Guo, and B.~Chen, ``{QED effect on a black hole
  shadow},'' \href{http://dx.doi.org/10.1103/PhysRevD.103.044057}{{\em Phys.
  Rev. D} {\bfseries 103} no.~4, (2021) 044057},
  \href{http://arxiv.org/abs/2012.07022}{{\ttfamily arXiv:2012.07022 [gr-qc]}}.

\bibitem{Zhong:2021mty}
Z.~Zhong, Z.~Hu, H.~Yan, M.~Guo, and B.~Chen, ``{QED effects on Kerr black hole
  shadows immersed in uniform magnetic fields},''
  \href{http://dx.doi.org/10.1103/PhysRevD.104.104028}{{\em Phys. Rev. D}
  {\bfseries 104} no.~10, (2021) 104028},
  \href{http://arxiv.org/abs/2108.06140}{{\ttfamily arXiv:2108.06140 [gr-qc]}}.

\bibitem{Liu:2024soc}
W.~Liu, D.~Wu, and J.~Wang, ``{Light rings and shadows of static black holes in
  effective quantum gravity},''
  \href{http://dx.doi.org/10.1016/j.physletb.2024.139052}{{\em Phys. Lett. B}
  {\bfseries 858} (2024) 139052},
  \href{http://arxiv.org/abs/2408.05569}{{\ttfamily arXiv:2408.05569 [gr-qc]}}.

\bibitem{He:2024amh}
K.-J. He, G.-P. Li, C.-Y. Yang, and X.-X. Zeng, ``{Observational features of
  the rotating Bardeen black hole surrounded by perfect fluid dark matter},''
  \href{http://arxiv.org/abs/2411.11680}{{\ttfamily arXiv:2411.11680
  [astro-ph.HE]}}.

\bibitem{Zhang:2022osx}
Z.~Zhang, H.~Yan, M.~Guo, and B.~Chen, ``{Shadows of Kerr black holes with a
  Gaussian-distributed plasma in the polar direction},''
  \href{http://dx.doi.org/10.1103/PhysRevD.107.024027}{{\em Phys. Rev. D}
  {\bfseries 107} no.~2, (2023) 024027},
  \href{http://arxiv.org/abs/2206.04430}{{\ttfamily arXiv:2206.04430 [gr-qc]}}.

\bibitem{Li:2024ctu}
G.-P. Li, H.-B. Zheng, K.-J. He, and Q.-Q. Jiang, ``{The shadow and
  observational images of the non-singular rotating black holes in loop quantum
  gravity},'' \href{http://arxiv.org/abs/2410.17295}{{\ttfamily
  arXiv:2410.17295 [gr-qc]}}.

\bibitem{Hou:2022eev}
Y.~Hou, Z.~Zhang, H.~Yan, M.~Guo, and B.~Chen, ``{Image of a Kerr-Melvin black
  hole with a thin accretion disk},''
  \href{http://dx.doi.org/10.1103/PhysRevD.106.064058}{{\em Phys. Rev. D}
  {\bfseries 106} no.~6, (2022) 064058},
  \href{http://arxiv.org/abs/2206.13744}{{\ttfamily arXiv:2206.13744 [gr-qc]}}.

\bibitem{Zhang:2024lsf}
Z.~Zhang, Y.~Hou, M.~Guo, and B.~Chen, ``{Imaging thick accretion disks and
  jets surrounding black holes},''
  \href{http://dx.doi.org/10.1088/1475-7516/2024/05/032}{{\em JCAP} {\bfseries
  05} (2024) 032}, \href{http://arxiv.org/abs/2401.14794}{{\ttfamily
  arXiv:2401.14794 [astro-ph.HE]}}.

\end{thebibliography}\endgroup
		
\end{document}